\documentclass[iop]{emulateapj}
\shortauthors{Sekanina}
\shorttitle{Outbursts and Fragmentation of Comet 168P}
\slugcomment{Version \today }
\begin{document}
\title{Temporal Correlation Between Outbursts and Fragmentation Events\\of
Comet 168P/Hergenrother}
\author{Zdenek Sekanina}
\affil{Jet Propulsion Laboratory, California Institute of Technology \\
  4800 Oak Grove Drive, Pasadena, CA 91109}
\email{Zdenek.Sekanina@jpl.nasa.gov}

\begin{abstract}
Outbursts are known to begin with a sudden appearance and steep brightening
of a\,``stellar nucleus'' --- an unresolved image of a plume of material on
its way from the comet's surface and an initial stage of an expanding halo of
ejecta.  Since the brightness of this feature is routinely reported, together
with astrometry, by most comet observers as the ``nuclear magnitude,'' it is
straightforward to determine the onset time, a fundamental parameter of any
outburst, by inspecting the chronological lists of such observations for a
major jump in the nuclear brightness.  Although it is inadmissible~to mix
nuclear magnitudes by different observers without first carefully examining
their compatibility, the time constraints obtained from data sets reported
by different observers can readily be combined.  The intersection of these
sets provides the tightest possible constraint on the outburst's onset time.
Applied to comet 168P/Hergenrother during its 2012 return to perihelion, three
outbursts were detected and their timing determined with good to excellent
accuracy.  Six fragmentation events experienced by the comet are shown to have
occurred in the same period of time as the outbursts.  Three companions~are
likely to have broken off from the primary in the first outburst, two
companions in the second outburst, and one companion in the last outburst.
All companions were short-lived, belonging to the class of excessively brittle
fragments.  Yet, the results suggest that most of the mass lost in the first
outburst remained relatively intact during the liftoff, while the opposite
was the case in the last outburst.
\end{abstract}


\section{Introduction}

Two classes of phenomena that always attract much attention of comet
observers are outbursts and nucleus' fragmentation.  Not unexpectedly,
they fairly often --- though not always --- correlate, but sometimes
it is not easy to provide convincing evidence for this correlation.
The solution is particularly difficult when there are several outbursts
and/or more than one companion to the primary nucleus in a relatively
short period of time.

The major observational difference between outbursts and fragmentation
(or splitting) events is that outbursts are detected practically as
they happen, subject only to (i)~the light time, (ii)~a large enough
amplitude for the observer to notice it as a jump in brightness, and
(iii)~his opportunity to observe the comet at the critical time.  By
contrast, sizable fragments of the nucleus, with separation velocities
in the submeter- to meter-per-second range, do not get resolved from the
parent nucleus until at least a few weeks, but more often a month or longer
after splitting.  Besides, nucleus' fragments cannot often be observed
without major interruptions because of their large, sudden brightness
fluctuations.  Finally, the rates of the fragments' relative motions are
temporally nonuniform, so that their separation times cannot be ascertained
by a linear extrapolation back in time.  This approximation invariably leads
to a grossly underestimated length of the interval between fragmentation
and observation.  The solution is likewise made more difficult by the
fragmentation hierarchy, in which a fragment of the first generation may
become the parent to a fragment of the second generation, etc.

With a high degree of confidence required to prove the relationship
between an outburst and a fragmentation event, it is necessary to
determine both the onset time of outbursts and the separation time
of fragments with high accuracy.  If there are several outbursts and
several fragments, the timeline of their hierarchy determines the degree
of accuracy with which the correspondence between two particular events
needs to be resolved.

Well determined onset times of outbursts are also critical in the
instances of inadequately observed companions, whose separation times
cannot be computed with much accuracy.  The outbursts' onset times are
then used to investigate the most probable correlations between outbursts
and fragmentation events.

\section{The Outbursts}

Any sudden, prominent, and unexpected brightening of a comet, caused
by an abrupt, short-term injection~of massive amounts of material from
the comet's nucleus into its atmosphere, is called an {\it outburst\/}.
By {\it sudden\/} is meant that the duration of its active stage usually~does
not exceed a fraction of the day or 1--2 days at~the~most.  The term
{\it prominent\/} conveys that the overall brightness increase during
the event is at least a factor of \mbox{2--2{\small $\frac{1}{2}$}}
(an amplitude of not less than 0.8--1.0 magnitude).  The word {\it
unexpected\/} implies that the event is not part of known periodic
of quasi-periodic variations, such as due to the nucleus' rotation.
Outbursts, especially the smaller ones, are frequent phenomena
experienced over centuries by a large number of comets, some of them
discovered while in outburst.  The propensity to such flare-ups
varies from comet to comet and it is not necessarily correlated with
heliocentric distance.  It is known that Comet 29P/Schwassmann-Wachmann
(Sec.\ 2.1), which never gets closer to the Sun than 5.4 AU, undergoes
major outbursts repeatedly, with an average rate of 7.4 events per year
(Trigo-Rodriguez et al.\ 2008, 2010).

Ordinary outbursts have some common features with the extremely rare
{\it giant (super-massive) explosions\/} (e.g., Sekanina 2008a), but in
other respects the two categories of phenomena differ from each other.
Ordinary outbursts originate, on the nucleus, from discrete active centers
(or regions) of a limited extent and always represent {\it local events\/}
on the scale of the nucleus' dimensions, with the total mass of ejecta
below ---  often by orders of magnitude --- 10$^{13}$\,grams (e.g.,
Sekanina 2008a).  Even though the mass released during an outburst
consists of both gas (active component) and dust, there is a wide range
of these events in terms of their mix.

\subsection{Historical Highlights:\ Dust Dominated and Gas Dominated
Outbursts}

While it is not the goal of this paper to examine possible production
mechanisms for cometary outbursts, descriptions of several well
documented historical examples of prominent explosive events and a
reference to some specific investigations of this subject are
relevant to the present objectives.

Critical aspects of explosive events in comets are related to the
formation of halos in the coma.  Reports of these features date back at
least to the early 19th century.  The first comprehensive report was
from Herschel (1847) in reference to the appearance of comet Halley in
late January 1836, observed by him from South Africa with his 47-cm
f/13 reflector.  Herschel wrote that ``\ldots the comet now was indeed
a most singular and remarkable object, \ldots a phenomenon, I believe,
quite unique in the history of comets.  Within the well-defined head
\ldots was seen a vividly luminous nucleus.''  Herschel's description of
what was the first observed case of a giant explosion (\mbox{Sekanina}
2008b) then continued with many additional details.  Most importantly
he commented on ``the extraordinary sharpness of termination'' of the
halo and was amazed that ``the comet was actually increasing in dimensions
with such rapidity that it might almost be said to be seen to grow!"
After the event the comet long remained anomalously bright, visible
with the naked eye for at least two months.  Although no spectroscopic
observations were made at the time, it is virtually certain that the
halo was composed of microscopic dust.

In 1883--1884 comet 12P/Pons-Brooks experienced a number of outbursts.
From a wealth of information in the literature, only a few reports are
mentioned below.  On one of the events, Struve (1884) in Pulkovo remarked
that on September 23, 1883, the comet looked like a star without any
nebulosity in a telescope finder but as a round, rather sharply bound
mass in the 38-cm refractor.  On the subsequent nights this halo grew
fainter, larger, and more diffuse and elongated.  After five days the
feature practically disappeared.  From the first three, most reliable
measurements of the expanding halo's dimensions reported by Struve, the
outburst should have nominally begun on September 22.94 UT, some 2 to
2{\small $\frac{2}{3}$} hours after the observations by \mbox{Schiaparelli}
(1883) and Abetti (1883), who both reported a very inconspicuous nucleus,
but about one hour before the comet was observed at Harvard by Chandler
(1883), who was ``astonished to find exactly in [the comet's] place a
bright, clearly defined star \ldots without sensible trace of nebulosity
\ldots that even an experienced observer would easily have failed to
distinguish \dots from \ldots stars''. Similarly, at about the same time
\mbox{Pickering} et al.\  (1900) commented: ``Comet resembles a star.
There has been a great change since yesterday.'' The next night
\mbox{Chandler} already saw the nucleus ``spread out into a confused,
bright disc with ill defined edges.''  The spectrum taken by
\mbox{Pickering} et al.\ on September 26 showed primarily the molecular
bands, with only a faint trace of continuum.  The event apparently
waned fairly rapidly.

Another major outburst of comet 12P, on January 1, 1884, was witnessed
at Potsdam by Vogel (1884) and by M\"{u}ller (1884a, 1884b).  Vogel
noticed a dramatic change in the appearance of the comet in a span of
two hours, during which a prominent, uniformly luminous, round disk
several arcseconds in diameter was formed.  Its spectrum was a pure
continuum.  The dimensions of the disk grew by 4$^{\prime\prime}\!$.1
in 33 minutes, from which the onset time on January 1.78 UT can be
calculated with an estimated uncertainty of not more than a fraction
of an hour.  This time was only about 1 hour before Vogel's second
observation.  The disk disappeared on the following days and the
continuous spectrum was then restricted only to a tiny nucleus.

M\"{u}ller's report is of great value, because the outburst occurred
literally before his eyes.  On January 1.77 UT he noticed that at the
location of the diffuse nucleus seen about 90 minutes earlier was now
``an almost perfectly point-like star \ldots at first sight so striking
\ldots [as if] a bright star was about to be occulted by the comet.''
M\"{u}ller's magnitude determination of the stellar nucleus with the use
of a Z\"{o}llner photometer indicated that it still grew in brightness,
reaching a fairly flat maximum around January 1.805 UT, then fading
gradually.  The overall evolution was so rapid that by January 1.90 UT
the feature already became distinctly less sharp.

The next extraordinary events were two episodes of a giant explosion
experienced by comet 17P/Holmes in 1892--1893.  The comet was actually
discovered in the course of the first episode, some 45--65 hours after
it had begun (as extrapolated from the rate of subsequent expansion;
Sekanina 2008a).  The spectroscopic observations made soon after the
discovery consistently showed the continuous spectrum to dominate, with
only a faint band sometimes reported mainly on the outside of the bright
disk or halo (Campbell 1893, Kammermann 1893, Vogel 1893).  The halo
continued to expand to gigantic dimensions, exceeding the Sun's diameter
about three weeks after the event's onset.  In small instruments the
comet's brightness was subsiding at a fairly slow rate, when the
beginning of a new explosion was detected by Palisa (1893) some 10 weeks
after the first one.  He reported (at an estimated 13--23 hours after the
onset of the explosion) that the comet looked like ``a yellow star, which
was surrounded by an envelope 20$^{\prime\prime}$ in diameter.''  The
envelope was a newly formed dust halo.  Numerous additional observers
provided similar accounts, with their summaries listed elsewhere (e.g.,
Bobrovnikoff 1943, Sekanina 2008a).

Comet 17P underwent an even more powerful giant explosion in October 2007,
when in a matter of about 2 days it brightened by an unprecedented 17
magnitudes (e.g., Sekanina 2008a) and was still observed with the naked
eye more than 4{\small $\frac{1}{2}$} months later!  The most conspicuous
feature of the post-peak branch of the light curve was a flat plateau,
with the total brightness (normalized to 1 AU from the earth) having
subsided by only 1 magnitude in the course of 4 months, as measured both by
the visual observers (e.g., Sekanina 2008a) and by a red-sensitive CCD
detector on the satellite Coriolis (Li et al.\ 2011).  The dust halo
was expanding at a rate of 0.5 km/s, losing gradually the symmetry and
reaching eventually the dimensions much greater than those of the Sun.

Bobrovnikoff (1932) became interested in the formation of halos in
comets after he investigated a number of such expanding features in the
head of \mbox{Halley's} comet (Bobrovnikoff 1931).  From spectroscopic
observations he concluded that they were of gaseous nature.  During
the 1986 apparition of Halley's comet, Schlosser et al.\ (1986) imaged
the evolution of 15 prominent CN halos (which they called shells) and
subsequently Schulz and Schlosser (1990) linked them to CN jets and
concluded that they both were made up of CHON particles.  Because these
features were not associated with a profound brightening of the comet,
their nature appears to differ from the halos seen in the early stage
of prominent outbursts.

The discovery of Comet 29P/Schwassmann-Wachmann in 1927 provided comet
astronomers with an object of unceasing propensity to outbursts, which
has ever since been subject for studying these phenomena.  The data
from the first 10--25 years of observation were summarized by Richter
(1941, 1954), who also compared the events in this comet with those in
other comets, including 12P/Pons-Brooks and 17P/Holmes (Richter 1949).
He concluded that the outbursts of different comets have some common
features and presented a timeline of an outburst, which can essentially
be summarized into six points:\\[-0.25cm]

(1) Before the outburst, the comet generally displays a diffuse coma
that sometimes is condensed toward the center and every now and then
exhibits a faint stellar nucleus.  The spectrum consists of molecular
bands.\\[-0.25cm]

(2) Within a time interval possibly as short as several minutes or
as long as an hour, the comet's appearance is being fundamentally
transformed.  A brilliant star (a preferable term would be a point-like
object --- {\it author\/}), which triggers a brightening by up to
8 magnitudes, appears in the center.  Its spectrum is continuous.
The former coma remains partially preserved during the outburst;
it may in part be outshined by the star, in part fade away.\\[-0.25cm]

(3) Shortly after the outburst, often only several hours later,
the stellar nucleus begins to grow steadily into a planet-like
disk.\\[-0.25cm]

(4) In the course of the next days the disk continues to grow.  The
comet's total brightness, which during this process has leveled off
or still risen, begins now to subside.\\[-0.25cm]

(5) After a few more days the comet regains its pre-outburst appearance
and so does its spectrum.\\[-0.25cm]

(6) The duration of these physical changes differs from case to case.
Even though  the course of events is the same, the scale of each outburst
entails a different time interval.

One may not agree with every detail of this description by Richter (1949),
but overall it does appear to recount the individual stages of evolution
of outbursts rather credibly.

In the same paper Richter also addressed the issue of expansion velocity,
finding values mostly on the order of hundreds of meters per second, and
he discussed a few possible production mechanisms.

In the decades since Richter's papers were written, countless numbers of
additional outbursts of comet 29P have been observed and studied.  An
excellent example is Beyer's (1962) account of a prominent outburst
in October 1959.  The dust halo was observed to expand for more than
30 days at a projected rate of 0.19 km/s, its maximum measurable
dimensions reaching almost those of the Sun.  The brightening, whose
initial rate was extremely steep, terminated about 4 days later, when
the comet reached an apparent visual magnitude 10.7.  Beyer's results
show that during the subsequent 30--40 days the light curve displayed
a flat plateau, with the brightness remaining essentially constant,
dropping by only 1 magnitude as late as 50--60 days after reaching
the peak.  It is obvious that this light curve is somewhat reminiscent
of that of the giant explosions, except that the flat plateau did not
extend for quite as long.  Even so, ejected dust with a long residence
time in the coma appears to dominate the outbursts of comet 29P, unlike
those of 12P.

The light curves of 29P/Schwassmann-Wachmann's outbursts published
by Trigo-Rodriguez et al.\ (2008, 2010) differ from that by Beyer (1962).
While the steep brightness jump at the outbursts' onset is as striking
as in Beyer's light curve, the peak appears much sharper, with the
brightness beginning to drop significantly only days afterwards.  This
effect is apparently due to the use by Trigo-Rodriguez et al.\ of a small,
10$^{\prime\prime}$ aperture, with which they sample only a fraction
of the coma to a distance, on the average, of some 40,000 km from the
nucleus.  Thus, Beyer's light curve is representative of the comet's
total brightness, Trigo-Rodriguez et al.'s light curves illustrate
brightness variations merely in the coma region nearer the nucleus.
It follows that with an adopted velocity of $\sim$0.2 km/s for the
coma expansion rate, the aperture covers only dust emissions less
than about 50 or 60 hours old.  By modeling a major outburst of 29P
in February 1981, Sekanina (1990, 1993) established from the feature's
morphology that its duration was about 0.7 the rotation period, or
3.5 days with \mbox{Whipple's} (1980) rotation period of 4.97 days.
Comparing the time scale of this outburst's evolution with its
morphology constraint, the rotation period could hardly exceed, or be
much shorter than, 5--6 days.  However, most values suggested in the
literature are in fact longer (Jewitt 1990, Stansberry et al.\ 2004,
Trigo-Rodriguez et al.\ 2010); on the other hand, Meech et al.\ (1993)
found a very rapid and complex rotation.

During the past decades, major outbursts have also been observed in a
large number of other comets, only a few of them being mentioned below.
Comet 41P/Tuttle-Giacobini-Kres\'ak underwent two enormous outbursts,
both with an amplitude of $\sim$9 magnitudes, 41 days apart during its
1973 apparition (Kres\'ak 1974).  Spectroscopic data showed that the
second outburst was dominated by molecular emissions (C$_2$, C$_3$,
CN, CH), with only a weak to medium-strength continuum present (Swings
and Vreux 1973).  However, from the similarities in the coma morphology,
duration (3 and 2 days, respectively; Kres\'ak 1974), light curve (the
rate of brightness subsidence only moderately gentler than the rate of
rise), and other attributes, it is likely that both outbursts were gas
dominated, resembling those of comet 12P.  The domination by gas in the
second outburst is also consistent with the absence of any major increase
in the brightness at close proximity of the nucleus and any sharply-bounded
halo around the time of the maximum total brightness; with the shrinking
of the bright coma from 110,000 km to 16,000 km in diameter between 2 and
3 days after the onset of the second outburst; and with the detection of
a {\it diffuse\/} nuclear condensation 3400 km in diameter at the first
of the two times (Kres\'ak 1974).  Comet 41P also experienced a rarely
mentioned postperihelion outburst during its 1995 return (Green 1995) and
three preperihelion outbursts within a span of about three weeks during
its 2000/2001 apparition (e.g., Sekanina 2008a).

Another previously faint periodic comet, 73P/Schwass\-mann-Wachmann,
entered its explosive era shortly before perihelion of its 1995 return,
when it underwent a 5-magnitude outburst, first detected --- on account
of the comet's proximity to the Sun in the sky --- with a radiotelescope
(Crovisier et al.\ 1996).  While no spectrum in the visible light is
available, it appears that no observable halo was formed during the
outburst, which accompanied a multiple fragmentation of the parent
nucleus (Sekanina 2005).  The comet's nuclear companions from 1995
continued to fragment during the fabulously favorable apparition of
2006 and probably also during the intervening return of 2000/2001, when
the comet was observed less extensively.

The complex correlation between nuclear fragmentation and outbursts was
exemplified by comet C/2001 A2 (LINEAR).  The parent nucleus --- also
called component B --- split, step by step, to generate six companions, A
and C--G, and underwent four outbursts, I--IV (\mbox{Sekanina} et al.\ 2002,
Jehin et al.\ 2002).  Outburst I coincided with the birth of companion A,
outburst II with companion C, and outburst III with companions D, E, and
F.  Outburst IV was not observed to correlate with any nuclear fragment,
while the birth of fragment G was not accompanied by any outburst.
According to Sekanina et al.\ (2002), a fragmentation event is or is not
accompanied by an observable outburst, depending on whether or not a
significant fraction of the fragment's mass disintegrates into dust upon
separation; and an outburst with no observed fragmentation event is the
outcome of the fragment's complete (or near-complete) disintegration.
These scenarios need to be kept in mind in the following investigation
of comet 168P.

From the wealth of information on exploding comets, it is concluded
that an outburst as such has no diagnostic significance for predicting
the future evolution of the object.  After undergoing an outburst, many
comets do not change their behavior at all.  For others, an outburst
triggers an extended period of enhanced activity, whereas for the
unfortunate few it portends their imminent cataclysmic demise.  Such
terminal flare-ups were exhibited, for example, by comets C/1999 S4
(LINEAR) and C/1996~Q1 (Tabur) shortly before their disintegration, but
the sequence of events observed in comet 168P is inconsistent with
a ``doomsday'' scenario.{\vspace*{-0.1cm}}

\begin{figure*}
\hspace*{-0.5cm}
\centerline{
\scalebox{0.679}{
\includegraphics{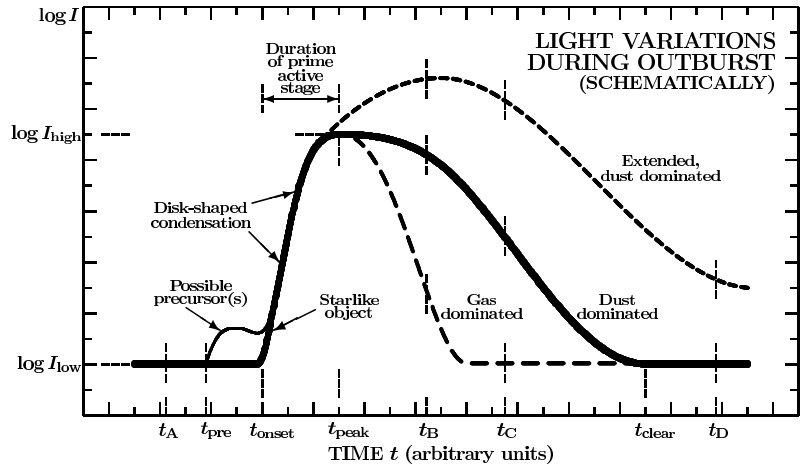}}} 
\caption{Schematic representation of an outburst.  The brightness, $I$,
is plotted on a logarithmic scale against time $t$.  Three categories
of events are depicted:\ a ``standard'' dust dominated (solid curve),
an extended dust dominated (short-dashed curve), and a gas dominated
(long-dashed curve).  All three events begin at the same onset time,
$t_{\rm onset}$, when the brightness is $I_{\rm low}$.  The precipitous
rise in the brightness, which includes the appearance of the starlike
object at the location of the nucleus and, later, the appearance of the
disk-shaped condensation, is terminated at time $t_{\rm peak}$, the end
of the prime active stage, when $I$ reaches a maximum, $I_{\rm high}$.
The brightness then begins to subside at a slower rate, until it drops
to the quiescent level $I_{\rm low}$ at time $t_{\rm clear}$.  By this
time, all material ejected during the outburst has left the volume of
coma photometrically investigated.  For an extended dust dominated
outburst, the coma continues to brighten after $t_{\rm peak}$ and its
brightness may remain elevated after $t_{\rm clear}$ because of a
persisting higher production rate of dust.  For the gas dominated
outburst, the brightness subsides more rapidly, reaching $I_{\rm low}$
long before $t_{\rm clear}$.  The main outburst may be preceded by
a minor precursor (thin curve), which starts at $t_{\rm pre}$.  The
symbol $I$ refers normally to the comet's total brightness, but it
could also apply to the brightness $I_N$ of the nuclear condensation.
The scenario is the same, but the rate of subsidence would then be
generally steeper.\\[0.2cm]}
\end{figure*}

The above examples of explosive phenomena in comets amply demonstrate
that observationally {\it each\/} such event begins exactly the same way, with
the appearance in the middle of the coma of a bright {\it starlike
object\/}, whose first sighting coincides with the onset of a {\it
precipitous rise\/} in the brightness of the central coma.  This
stellar feature is an initial stage of an expanding halo (or disk)
of material, whose surface brightness gradually decreases until its
eventual disappearance, while its integrated brightness may for a while
continue to increase with time, depending on the amount of released
material, on the relative contributions from gas and dust, on the
size-distribution of dust particles, and on the post-outburst physical
conditions in the active region from which the outburst originated.
When the outburst is dominated by gas, the rate at which the brightness
generally subsides is determined primarily by the photodissociation
lifetime of the observed molecules (such as C$_2$, CN, etc., in the
visible light), which does not exceed a day or two near 1 AU from
the Sun.  However, these photodissociation products, contributing
substantially to the brightness of the outer coma,  do not have any
effect on the region of nuclear condensation, where ejected dust
appears to prevail even in the gas dominated outbursts.  Because of
smaller amounts of dust involved, the post-peak comet brightness in
these events drops more rapidly with time and the expanding disk
disappears soon.  By contrast, when the outburst is dominated by dust,
the brightness subsides more gradually and the expanding dust halo,
while changing its morphology, survives longer.\footnote{The dust
halos originating in the giant explosions survive by far the longest.}
If the outburst triggers an episode of continuing dust emission from
the source or nearby areas on the nucleus, the brightness may remain
elevated for an extended period of time.  Finally, the shape of the
light curve also depends on the comet's position in the orbit
(preperihelion vs.\ postperihelion, heliocentric distance, etc.) and
on the diurnal and/or seasonal activity variations at the location of
the emission region.

\subsection{Outbursts and the Nuclear Magnitudes}

The purpose of this paper is to convince the reader that CCD data
sets of the nuclear-condensation brightness (not to be confused with
the true brightness of the comet's nucleus), routinely reported in
terms of the so-called {\it nuclear magnitudes\/} to the {\it IAU Minor
Planet Center (MPC)} as part of astrometric observations, can serve
as the basis to a simple, straightforward technique for efficiently
constraining the onset time of outbursts.

Given the poor reputation of {\it reported nuclear magnitudes\/}, this
statement appears at first sight to be nothing short of heresy.  Indeed,
in smaller telescopes the nucleus is always hidden in a much brighter
condensation that surrounds it, and the observer is in no position to
rectify the problem.  It gets so bad that, for example, the glossary of
the {\it International Comet Quarterly (ICQ)\/},\footnote{Consult the
subject items ``{\sf m}$_2$'' and ``{\sf Magnitude}'' in the ICQ web site
{\tt http://www.icq.eps.harvard.edu/ICQGlossary.html}.} emphasizes that
these magnitudes are ``fraught with problems \ldots especially because
[they] are extremely dependent upon instrumentation \ldots and
wavelength.  Nuclear magnitudes are chiefly used for astrometric
purposes, in which predictions are made for the brightness of the
comet's nuclear condensation so that astrometrists can gauge how faint
the condensation is likely to be and thus how long an exposure is
needed to get a good, measurable image \ldots [of] the site of the
main mass of any comet.''

As also mentioned in the ICQ glossary, the nuclear magnitudes of
comets used to be designated as $m_2$ in the ephemerides of comets, but
``in 2003 a subcommittee of IAU Commission 20 \ldots decided that the
concept of `nuclear' magnitudes should be done away with \ldots [and]
since then the heading `{\sf Mag.}' \ldots refer[s] to the predicted
brightness of comets.''  Whereas comet ephemerides no longer provide
{\it predicted\/} values of nuclear magnitudes, the MPC's report format
for the optical astrometric observations of comets to be submitted for
publication in the {\it Minor Planet Circulars\/} and the {\it Minor
Planet Electronic Circulars\/}\footnote{See the information web site
of the {\it IAU Minor Planet Center\/}
{\tt http://www.minorplanetcenter.net/iau/info/OpticalObs.html.}}
continues to allow one to list the nuclear magnitudes with a flag
``{\sf N}'' (as opposed to ``{\sf T}'' for the ``total'' magnitudes)
in column 71.  A great majority of comet observers has indeed to this
day been providing the nuclear magnitudes of comets in this fashion.

The sudden appearance, at the location of the nucleus, of the starlike
object signals the beginning of release from the surface of a major plume
of material, activated by a surge of erupting gases from the underneath.
Measured with a small sampling aperture, the nuclear magnitude is much
more sensitive to both the initial starlike stage of the outburst and to
the steep brightening of the expanding plume (that is, the halo) than
is the comet's total magnitude.  Hence, the same property of the nuclear
magnitudes that makes them unattractive for other scientific studies is
now deliberately exploited.  To my knowledge, this approach has never
been employed before.  In practice, caution need be exercised in examining
the published information, because the {\it nuclear magnitude\/} $H_N$
--- the quantity used to characterize the brightness $I_N$ of the nuclear
condensation --- may, as already pointed out, vary from observer to
observer.  Two caveats deserve particular attention:

(1) It is inadmissible to combine sets of nuclear magnitudes $H_N$,
reported by different observers, unless they are {\it proven\/} to
be compatible by careful analysis; and

(2) The detection of an outburst can only be regarded as secure, if
the timing of its onset is consistently and independently confirmed
by all, or at least an overwhelming majority of, the relevant sets of
nuclear-magnitude data reported by the observers during the critical
period of time.

On the other hand, a great advantage of this approach is the fact that
information on the nuclear-condensation brightness is listed by nearly
all observers who report their astrometric results.  Accordingly, for
most comets, including 168P, extensive sets of CCD nuclear magnitudes
are available for application of this technique.

\subsection{Temporal Photometric Profile of an Outburst}

Schematically, the brightness variations during an outburst are expected
to follow one (or be a combination) of the light curves in Figure 1.
The solid curve is a generic brightness profile for a dust dominated
outburst.  The event starts at time $t_{\rm onset}$, when the bright
stellar object first appears at the location of the nucleus and the
sheer brightness rise begins.  The starlike feature is the initial
stage of an expanding luminous disk, whose brightness peaks at $t_{\rm
peak}$.  The quantity \mbox{$2.5 \log(I_{\rm high}/I_{\rm low})$} is
the amplitude of the outburst in magnitudes, whereas the interval
\mbox{$t_{\rm peak} - t_{\rm onset}$} is the duration of its prime
active stage, assuming that it is shorter than the residence time of
dust particles within the measured boundaries of the coma and that the
luminous disk is optically thin (as is almost always the case even
near the nucleus itself), and in the absence of dust fragmentation.  At
$t_{\rm peak}$ the brightness begins to subside, first very slowly,
until it eventually drops to the quiescent level $I_{\rm low}$ at
time $t_{\rm clear}$.  By this time, the withdrawal from the volume of
the coma of all material ejected during the outburst is completed.
With the next event, the whole cycle is repeated.

The brightness of an extended dust dominated outburst continues to rise
after $t_{\rm peak}$ because of a persisting higher production rate of
dust (or for another reason, such as dust particle fragmentation).  For
example, the prime event may be followed by secondary outbursts (caused,
e.g., by impacts of boulders in ballistic trajectories back on the
surface, thus opening new emission centers), which in some cases could
lead to more or less permanently elevated activity, continuing to fill
the coma with large amounts of new dust.

In the gas dominated outbursts the post-peak brightness subsides more
rapidly than in the dust dominated outbursts, reaching $I_{\rm low}$
long before $t_{\rm clear}$.  The halo, containing dust, disappears
soon after the outburst's onset.

The prime active stage of any outburst may be preceded by a precursor,
a minor flare-up that indicates that the main event is in the making.
Since an outburst is essentially the product of succumbing to a stress
applied to the surface at a particular location of the nucleus, the
precursor could very well be the sign of the nucleus' limited initial
resistance to the straining force.

In this subsection, the brightness $I$ --- as well as $I_{\rm low}$
and $I_{\rm high}$ --- has been understood to refer to the coma, or,
more precisely, to the coma within its measured boundaries.  Since
there are no constraints on the boundaries, $I$ is, generically, the
brightness in any volume of the coma centered on the nucleus, and may
therefore also indicate $I_N$ [and similarly $(I_N)_{\rm low}$ and
$(I_N)_{\rm high}$], the brightness of the nuclear condensation, as
%
%
derived, in terms of the nuclear magnitude $H_N$, from the measurements
of the CCD images through a small sampling aperture.  

\subsection{Method for Constraining the Onset Time of\\an Outburst
from Sets of Nuclear Magnitudes}

I now consider a dust dominated outburst (solid curve in Figure 1)
and a set of nuclear-brightness data, $(I_N)_j$ \mbox{($j = 1, 2,
\ldots$)}, reported by a particular observer.  Let the first $k$
observations be made before the outburst's onset, so that at any
time $t_j$ \mbox{($j = 1, 2, \ldots, k$)} that satisfies a condition
\mbox{$t_j < t_{\rm onset}$}, such as $t_{\rm A}$ in Figure 1, the
expected nuclear brightness is \mbox{$(I_N)_j \simeq (I_N)_{\rm
low}$}.  Let the next \mbox{$n-k$} observations be made, by the same
observer with the identical telescope, after the outburst's onset,
but before all dust ejecta evacuate the region of the nuclear
condensation whose brightness the observer measures.  These times
satisfy a condition \mbox{$t_{\rm onset} < t_j < t_{\rm clear}$}
\mbox{($j = k+1, k+2, \ldots, n$)}, such as $t_{\rm B}$ or $t_{\rm
C}$ in Figure 1, and the brightness is then \mbox{$(I_N)_{\rm low}
< (I_N)_j \leq (I_N)_{\rm high}$}.  Perfunctory inspection of the
set of nuclear-brightness data usually suffices to detect the sudden
jump in $I_N$ between times $t_k$ and $t_{k+1}$ and to conclude that
the outburst began at some point between the two times,
\begin{equation}
t_k < t_{\rm onset} < t_{k+1}.
\end{equation}
This result, derived from the particular observer's data, formally
provides the expressions for a probable time of the outburst's onset,
\mbox{$\langle t_{\rm onset} \rangle = \mbox{\small $\frac{1}{2}$}
(t_k + t_{k+1})$}, and its uncertainty, which is equal to \mbox{$\pm
\mbox{\small $\frac{1}{2}$} (t_{k+1} - t_k)$}.  It is noted that
no information on the outburst can be extracted from observations
made at times \mbox{$t_j > t_{\rm clear}$}, that is, at
\mbox{$j > n$}, such as at $t_{\rm D}$ in Figure 1.  If no
observation has been made between $t_k$ and $t_{n+1}$, that is,
when \mbox{$n = k$}, the observer has missed the outburst.

Next, I consider a total of $\nu$ observers that provide information on
the comet's nuclear brightness before and during the outburst.  Let
the brightness data by an $i$th observer \mbox{($i = 1, \ldots, \nu$)}
constrain, in analogy to condition (1), the outburst's onset time to
an interval \mbox{$t_i^- < t_{\rm onset} < t_i^+$}, and let the set of
all times between $t_i^-$ and $t_i^+$ be called {\bf A}$_i$,
\begin{equation}
\mbox{\bf A}_i = (t_i^-\!,\, t_i^+),
\end{equation}
where the parentheses mean an open interval, with the boundaries
excluded.  The resulting constraint, obtained by combining those
from the data by all $\nu$ observers, is then represented by the
intersection {\bf A} of the sets {\bf A}$_i$,
\begin{eqnarray}
\mbox{\bf A}\, & = & \bigcap_{\,i=1}^{\,\nu} \! \mbox{\bf A}_i = \left(
  \max[\,t_1^-\!, t_2^-\!, \ldots, t_\nu^-], \min[\,t_1^+\!, t_2^+\!,
  \ldots, t_\nu^+] \right). \nonumber \\[-0.4cm]
 & &
\end{eqnarray}
Thus, while the brightness data by the individual observers should not
be mixed, the temporal constraints derived from them can readily be
combined.

Valid constraints can be obtained even from the sets of nuclear-brightness
data by the observers who saw the comet only before $t_{\rm onset}$ or
only after $t_{\rm onset}$ (but before $t_{\rm clear}$, of course), once
one knows the tentative constraints on the onset time from the data sets
by other observers.  If all of the brightness data reported by an observer
$p$ \mbox{($p \leq \nu$)} at times close to this range are near his own
$(I_N)_{\rm low}$ value, then his last observstion, made at time $t_p^-$,
can be incorporated into condition (3) as a valid constraint.  Similarly,
if all of the brightness data reported by an observer $q$ \mbox{($q \leq
\nu$)} at times close to this range are much greater than his own
$(I_N)_{\rm low}$ value, then his first observation, made at time $t_q^+$,
can likewise be incorporated into condition (3) as a valid constraint.
On the other hand, the times $t_p^+$ and $t_q^-$, referring to these
observers' missing brightness constraints at the other end of the time
interval, are obviously indeterminate, can be put equal to, e.g.,
\mbox{$t_p^+ \rightarrow {\textstyle +\infty}$} and \mbox{$t_q^-
\rightarrow {\textstyle -\infty}$}, and have no effect on the condition
(3).  The expression for the probable onset time of the outburst and its
uncertainty resulting from the applied set of constraints is finally
\begin{eqnarray}
\langle t_{\rm onset} \rangle & = & {\textstyle \frac{1}{2}} \! \left\{
  \max[\,t_1^-\!, t_2^-\!, \ldots, t_\nu^-] + \min[\,t_1^+\!, t_2^+\!,
  \ldots, t_\nu^+] \right\} \nonumber \\[-0.08cm]
 & & \llap{$\pm$} {\textstyle \frac{1}{2}} \! \left\{ \min[\,t_1^+\!,
  t_2^+\!, \ldots, t_\nu^+] - \max[\,t_1^-\!, t_2^-\!, \ldots, t_\nu^-]
  \right\}, \nonumber \\[-0.1cm]
 & & 
\end{eqnarray} \\[-0.45cm]
%
where $\min[\,t_1^+\!, \ldots \,] \!>\! \max[\,t_1^-\!, \ldots \,]$.  This concludes
the exercise.

\begin{table*}
\centerline{
\scalebox{0.78}{
\includegraphics{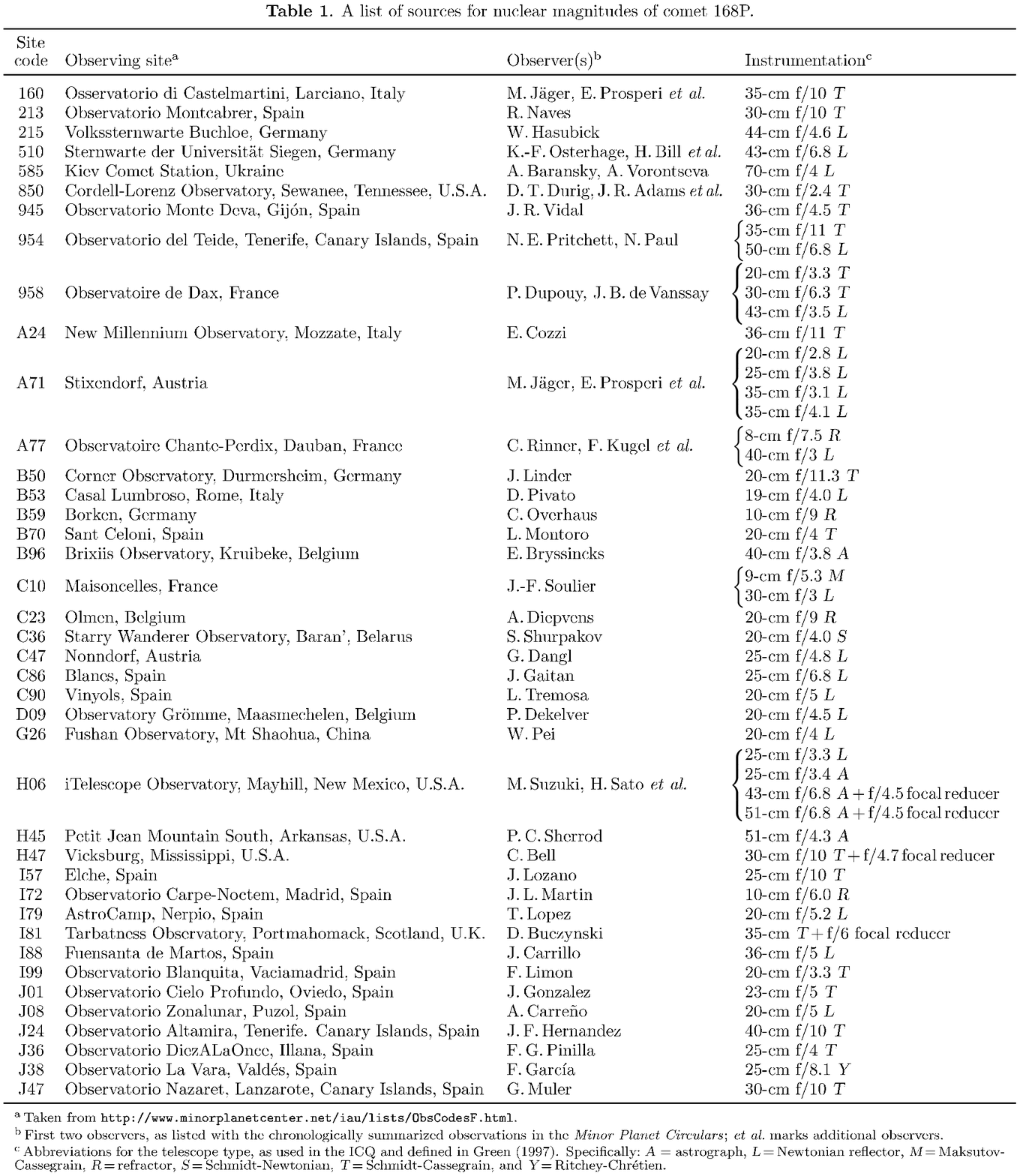}}} 
\vspace*{-1.4cm}
\end{table*}

\section{Comet 168P/Hergenrother in 2012}

Discovered by C. W. Hergenrother in November 1998 on images taken by
T. B.  Spahr, this short-period comet with a perihelion distance of
1.4 AU remained very faint during its observed returns to perihelion
in 1998 and 2005.\footnote{See, e.g., {\tt
http://cometography.com/pcomets/168p.html.}}  In 2012, an extremely
favorable return, its apparent magnitude was expected to reach 15 near
perihelion, which occurred on October 1.

The first published indication of a major deviation from the expected
evolution was a visual observation by Gonzalez (2012), who reported
the comet to be of magnitude 11.2 in his 20-cm reflector on September
6.90~UT.  The comet then continued to brighten, reaching a total
magnitude of at least 9 during October (e.g., Green 2012).  The
comet was more than 4 magnitudes brighter than expected in early
September and at least 6 magnitudes brighter than expected during
October.\footnote{The predicted magnitudes are at the {\it Cometary
Science Laboratory\/}'s site:\ {\tt
http://www.csc.eps.harvard.edu/168P/index.html.}}

\subsection{The Outbursts of Comet 168P}

Applying the described method, I was able to detect not one, but three
consecutive outbursts of this comet in a time span of one month.  The
search began by collecting the sets of nuclear magnitudes reported to
the MPC by the astrometric observers from 40 locations (Spahr et al.\
2012).  Information on these observing sites is summarized in Table 1,
the individual columns listing successively:\ the IAU site code (as
assigned by the MPC), the observatory's name and/or location, the
name(s) of the observer(s), and the instrumentation used.

The sets of input data for the outburst search are presented in Tables
2--5.  Tables 2--4 have identical format and list the data sets
relevant to, respectively, outbursts I, II, and III.  In each of these
tables the data are arranged by the observatory in column 1, with the
dates of observation, $t_{\rm obs}$, following in column 2 chronologically.
 More specifically, because it is customary to take several images
during each night, it is the interval from the mid-exposure time of
the first image to the mid-exposure time of the last image that is
listed to 0.001 of a day.  This interval usually amounts to a fraction
of one hour, but there are exceptions, with a longer span sometimes
covered.  Occasionally, long sequences of images were taken, in which
case more data sets are tabulated for the same date.  Each interval is
then converted to a range of times reckoned from the comet's perihelion
time, $t_\pi$, in column 3.  Columns 4, 5, and 6 show, respectively, the
nuclear magnitude $\langle H_N \rangle$, calculated as an average of the
magnitudes from the individual images provided by the observer(s), its
root-mean-square error, and the number of images reported.  The same
nuclear magnitude from all images is marked by a dash in column 5.  The
final column, one line per observing site, provides the constraint, in
terms of a range of allowed onset times, resulting from the sets of
images taken at the given observing site and reckoned again from the
perihelion time.  The relationship of these constraints to the terms
used in the method from Sec.\ 2.4 is discussed in the next paragraph.
In addition, each outburst is described by an {\it average magnitude
jump\/}, calculated from the tabulated differences in $\langle H_N
\rangle$ that bracket the derived onset time.  This quantity is a crude
measure, on the magnitude scale, of the outburst's perceptibility.  It
is useful for assessing a level of one's confidence that the outburst
occurred (the greater the jump, the more confident one feels), but does
not characterize its strength and has no direct relationship to the
amount or mass of the material ejected in the event.

If the entry in {\vspace*{-0.01cm}}column 7 consists of two numbers,
they indicate, respectively, {\vspace*{-0.06cm}}times \mbox{$t_i^-
- t_\pi$} and \mbox{$t_i^+ - t_\pi$}, where $t_i^-$ and $t_i^+$ are
the boundaries of the set {\bf A}$_i$ in equation (2) and $t_\pi$
is the comet's perihelion time.  For example, in the data set from
observing site A71 for outburst I in Table 2, the images from the
first three dates --- on August 15, August 28, and September 9 ---
give the average nuclear magnitudes of 16.9, 17.5, and 15.5, each
with an uncertainty of $\pm$0.1 magnitude.  According to the observers
at this site, the nuclear condensation apparently faded a little
between the first two dates, with no evidence of an outburst prior to,
and including, August 28.039~UT.  However, the brightness jumped up by
fully 2 magnitudes between that time and September 9.865 UT, the time
of the first image on the 9th, so the outburst must have occurred in
the intervening period of time.  This is consistent with Gonzalez's
(2012) observation mentioned above.  When reckoned from the perihelion
time, August 28.039 UT is equivalent to $-$34.933 days, while September
9.865 UT becomes $-$22.107 days, which are indeed the two entries
listed as the boundary constraints for the onset of outburst I in the
last column of Table 2 from the nuclear magnitudes provided by site A71.
To call the reader's attention to the magnitude jump, the entries in
column 7 are positioned between the rows of the two boundary dates and,
in addition, a wedge separates these two rows in the nuclear-magnitude
column.

If the data reported in Table 2 by observing site A71 were the only
constraint available, the probable onset time of outburst I, would
have been, following (4) and after rounding off, \mbox{$\langle
t_{\rm onset} \rangle = {\rm September} \; 3.5 \pm 6.4$} UT, or
\mbox{$\langle t_{\rm onset} \rangle - t_\pi = -28.5 \pm 6.4$} days.
Table 2 shows, however, that there is a total of 12 constraints, which
narrow down the uncertainty considerably and offer for the onset time
the tightest limits, which are shown by the entries in the slanted
type style in column 7:\ the maximum value of $t_i^-$ comes from
observing site 958, the minimum value of $t_i^+$ from site C86.  The
result, in Table 6, shows that the outburst began most probably on
September 1, two days earlier than indicated above by the constraints
from site A71, and that the uncertainty is more than 4\,times smaller.
The average magnitude jump from the 9 two-sided constraints is
\mbox{$1.7 \pm 0.6$} magnitudes, and the first detection of the
outburst by Gonzalez (2012) apparently occurred between about 4 to
7 days after it had begun.

The results reported by observing sites C86, D09, and J01 are examples
of the post-outburst observations that could be incorporated into
Table 2 as further constraints on outburst I, because in each case the
nuclear condensation was fading within enough time (5--12 days) after
the event.  Of these, C86 was in fact instrumental in reducing the
error of the result, because no other observations were made on
September 3 and the comet was not observed at all on September 1, 2,
and 4.\footnote{See the list of astrometric observations of 168P in
the MPC database on {\tt http://www.minorplanetcenter.net/db\_search}.}

Outburst II, for which the input data are summarized in Table 3, differed
from outburst I in that it clearly had a precursor.  The total number of
constraints on the timing of the main event equals 14, described again
in column 7.  The maximum value of $t_i^-$ comes from site H47 and the
minimum value of $t_i^+$ from site C47.  For each of the 14 sites, a
large wedge marks the outburst in the column for the nuclear magnitudes.
From sites H47 and I81 the comet was observed only before the outburst's
onset, from site I57 only after it.  From the 11 two-sided constraints
the average magnitude jump equals \mbox{$2.4 \pm 0.6$} magnitudes; the
resulting onset time is in Table 6.

The precursor to outburst II appears in six of the 14 data sets in
Table 3, from sites 958, C36, C47, I72, I81, and J24.  The precursor's
constraints, not listed in Table 6, are marked by the small wedges in
column 4 of Table 3.  It appears that the precursor began most probably
just before September 20.0 UT, more than 2 days before did the main
event.  The precursor does not show up distinctly in the nuclear
magnitudes from C23 and it is not detected in the magnitudes from the
other sites, in part because of their unfavorable timelines.  The
data from sites H47 and J24 suggest that elevated activity culminating
in outburst II may have begun even before September 19 (small wedges with
a question mark), but this is not supported by the other tabulated data.
In any case, there is no doubt that the dust emission rate during much
of September was increasing first gradually, before eventually erupting
in outburst II.  From the six detections of the precursor, its resulting
magnitude jump is found to be, on the average, \mbox{$0.7 \pm 0.2$}
magnitude.

\begin{table*}
\centerline{
 \scalebox{0.76}{
 \includegraphics{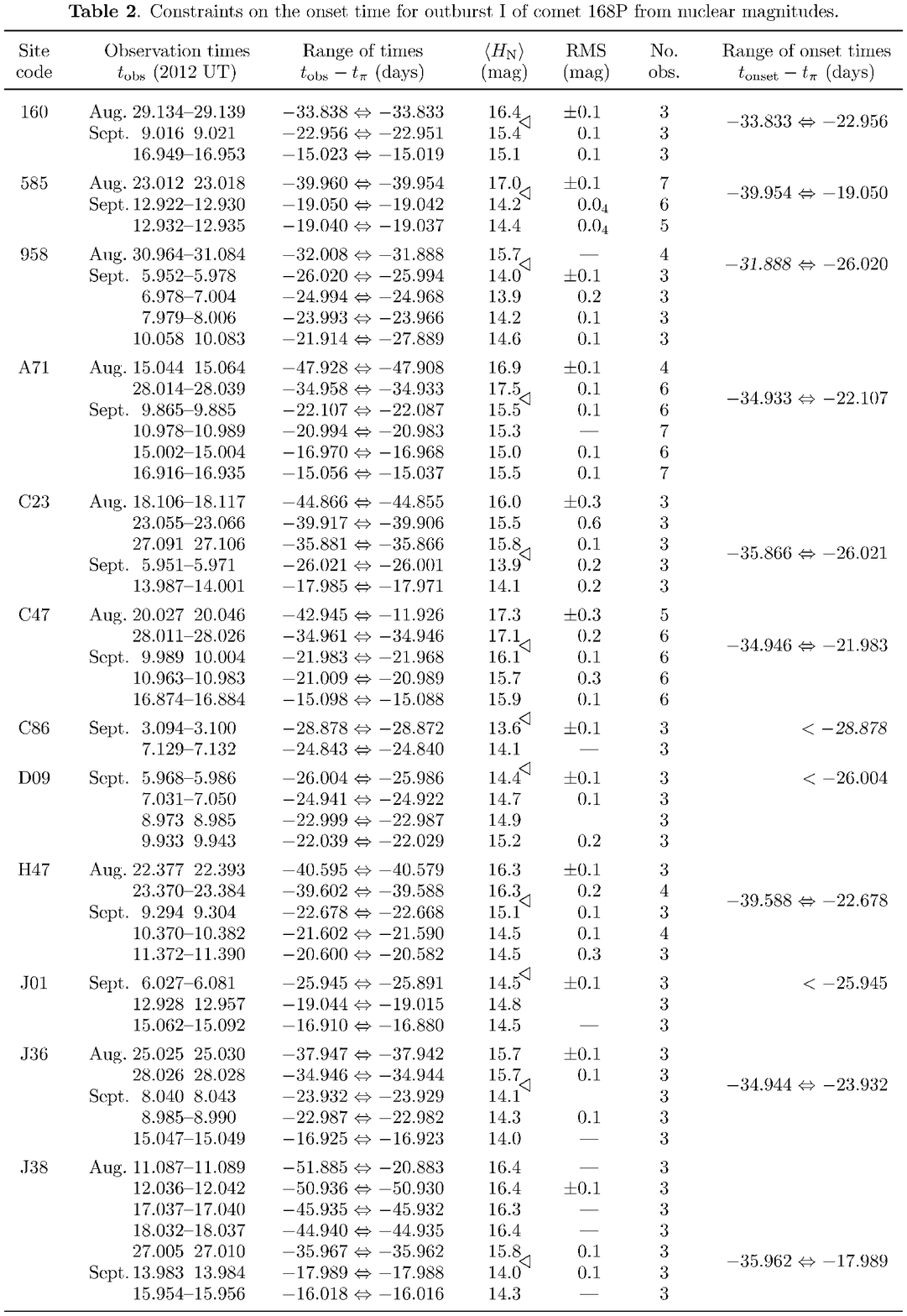}}} 
\vspace*{0.5cm}
\end{table*}

\begin{table*}
\centerline{
 \scalebox{0.73}{
 \includegraphics{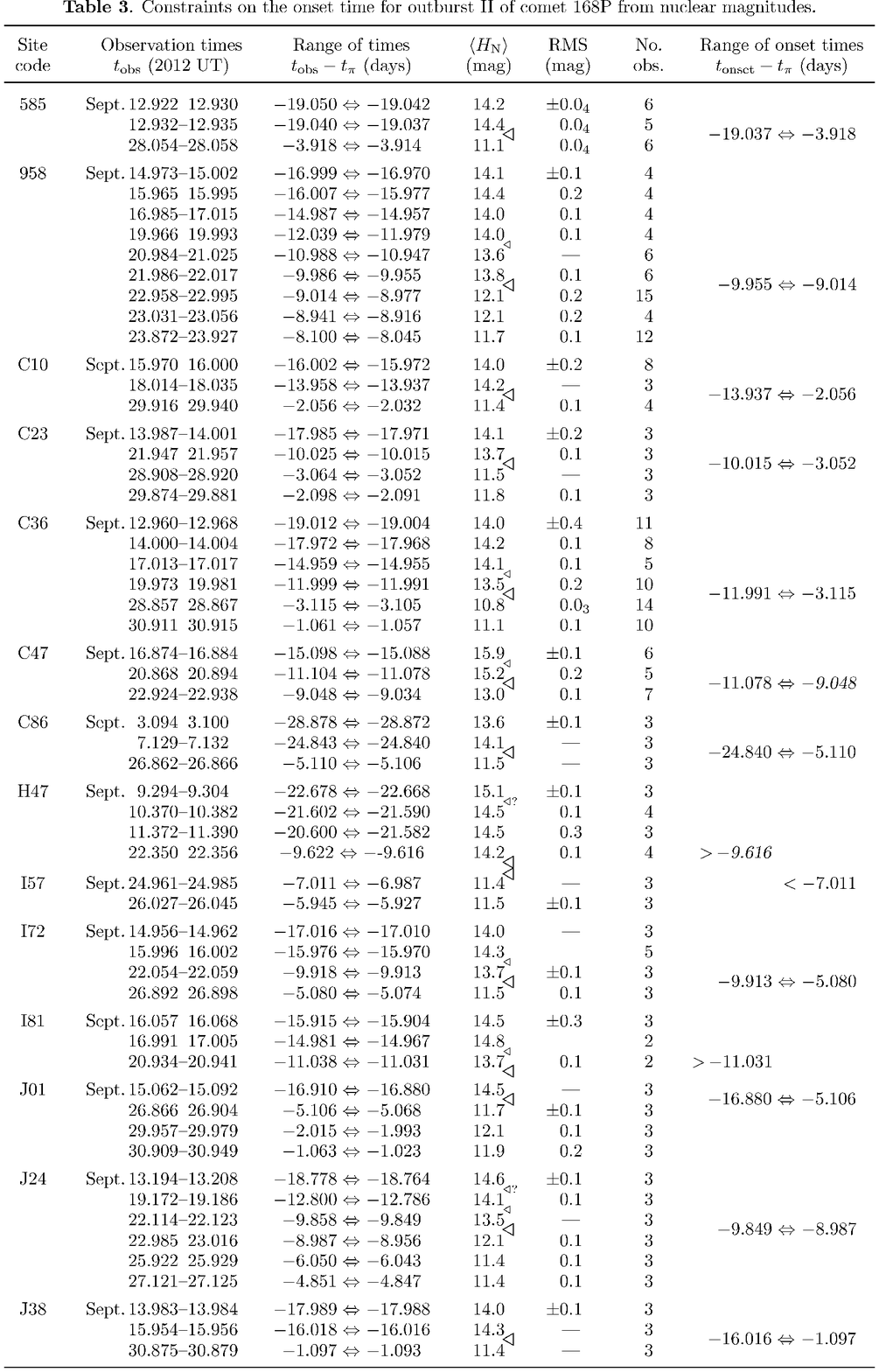}}} 
 \vspace*{0.5cm}
\end{table*}

Outburst III is the most difficult test of the proposed technique for
detecting the timing of these events, because it has by far the smallest
amplitude of the three.  The relevant data set is in Table 4, which
presents the nuclear magnitudes from 13 observing sites.  The comet
was observed only before the event from site C23 and only after the
event from sites 213 and I99.  The data from the remaining 10 sites
bracket the onset time of the outburst, but two of these sites failed
to register it, as discussed later in this paragraph.  From the eight
remaining constraints, the maximum value of $t_i^-${\vspace*{-0.05cm}}
amounts to October 1.80 (site A77) and the minimum value of $t_i^+$,
October 1.78 (site I57).  This result{\vspace*{-0.05cm}} is in conflict,
albeit marginal, with the condition \mbox{$\max[\,t_i^-\!, \ldots\,] <
\min[\,t_i^+\!, \ldots\,]$} mentioned below expression (4).  Table 4 shows
that the observing session at site A77 completely overlapped the shorter
session at site I57, and in both cases the reported magnitude jump was
only 0.4 magnitude.  Most importantly, the magnitudes reported from I57
are fainter than those from A77, so that the sampling aperture used at
I57 was probably smaller and the reported nuclear magnitudes are more
diagnostic of the innermost-coma region and of the plume of material
leaving the surface of the nucleus.  Therefore, as listed in Table 6,
outburst III must have begun \mbox{during}, or just moments before, the
observing session at site I57, and the onset time is determined with
accuracy better than $\pm$0.1 day.  The minor discrepancy between the
constraints from sites I57 and A77 illustrates that the recognition of
an outburst's onset depends, to a degree, on the details of imaging
observations (Sec.\ 3.2).  As already mentioned, outburst III was not
detected at two of the 13 sites, J08 and J24, even though in both cases
the observations do bracket the onset time established by the data from
the other sites (Table 4).  Closer inspection shows a 7-day gap between
the two J08 entries that bracket outburst III, the first having been made
during, or shortly after, outburst II.  Similarly, the second of the two
J24 observations that bracket outburst III was made on October 6, more than
4 days after the event's onset.  These cases illustrate the advantage of
having a dense timeline.  Indeed, every site that provided consistent
constraints featured at least one observation from the time span of
October 1--4.

\begin{table*}
\vspace*{-0.65cm}
\centerline{
 \scalebox{0.825}{
 \includegraphics{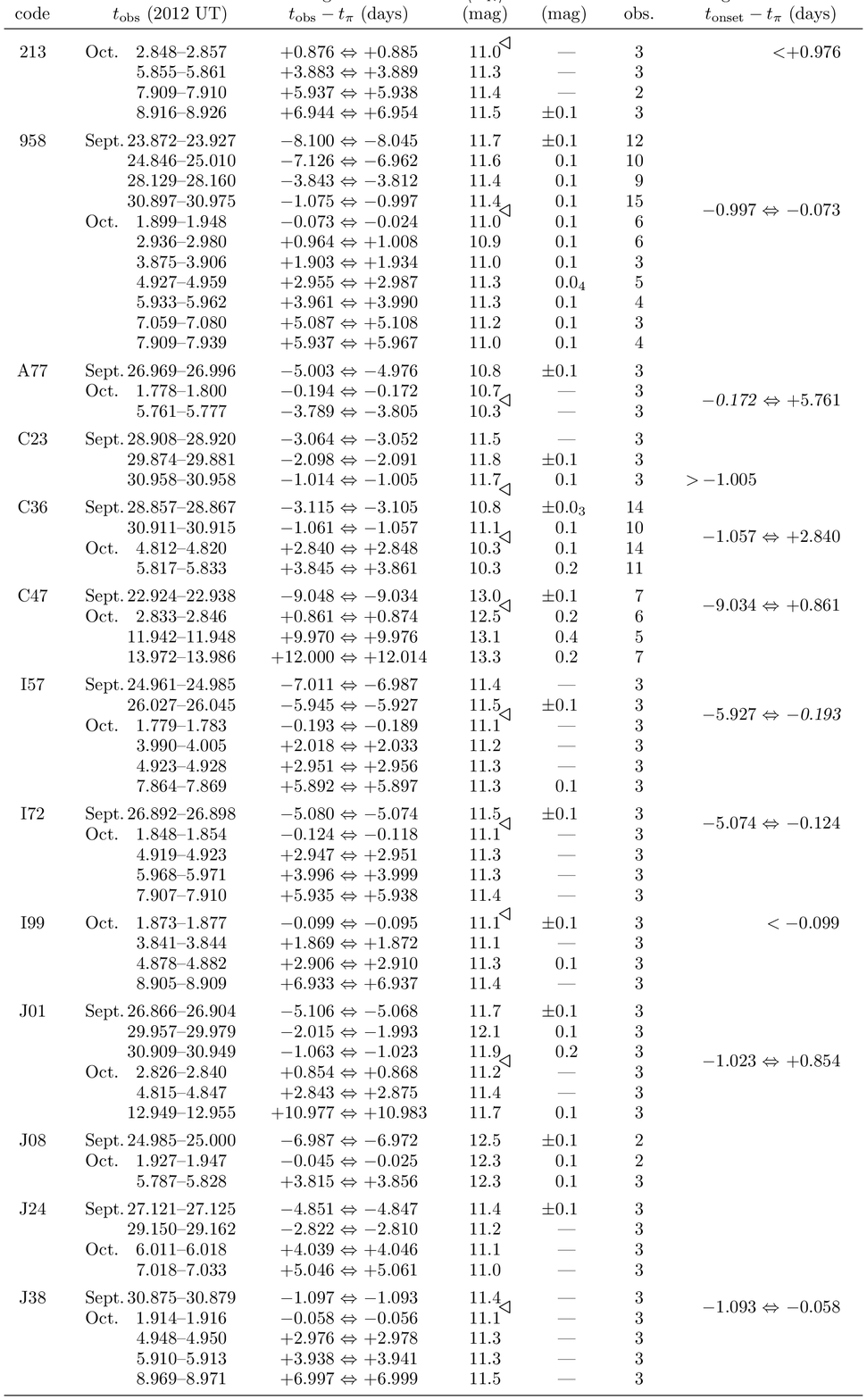}}} 
\end{table*}

The data from the eight sites that did constrain the onset of
outburst III from both sides were also used to compute the
average magnitude jump in this event, which was found to be
\mbox{$0.5 \pm 0.2$} magnitude.  As a fair lower limit to the
event's amplitude, this value suggests that the October 1
flare-up was probably barely what was accepted in Sec.\ 2 as a
minimum brightening that still deserves to be called an outburst
(an amplitude of 0.8--1.0 magnitude).  If so, it is nothing short
of remarkable that the method of nuclear magnitudes turned out to
be as successful in detecting outburst III as the above account
demonstrates.

The continuing search along the near-perihelion orbital arc in a
massive data set starting in early October, several days after
the onset time of outburst III, revealed no further explosive
events.  Thus, one of the primary tasks of this investigation has
been completed.

For the data from the post-outburst period of comet 168P in Table
5, the listed six columns are identical to the first six columns
in Tables 2--4.  In the absence of further outbursts, the seventh
column is not in Table 5 needed.  Out of the total of 34 observing
sites included in Table 5, the nuclear magnitudes from 18 --- 213,
215, 945, 954, A24, B50, B53, B59, B70, B96, C36, C47, G26, I57, I72,
I88, J01, and J24 --- show, within the errors involved, no clear sign
of deviation from an essentially continuous, even though somewhat
uneven, brightness decrease with time during the entire period
from the first week of October until December 11, when this study
of the comet's activity is terminated.  On the other hand, the
data from the 16 remaining sites do show one or more instances of
temporally localized brightening.  These potential events are
marked in Table 5 by wedges with a question mark.  The existence
of some of them appears to be supported by the data from more than
one site.  Fully 13 of the 16 sites --- 510, 958, A71, A77, C10,
C23, C90, H06, H47, I79, I99, J38, and J47 --- show at least one
episode of brightening in the broad time span between October 21
and November 7.  Two of these sites suggest more such episodes:\
site 958 implies two pairs of them, the first pair between October
23 and 27 and between October 27 and 28, the second pair between
November 3 and 5 and between November 5 and 6.  Site A77 indicates
two episodes, one between October 23 and 29 and the second between
October 29 and November 2.  Yet, the data from sites 213, 215, 850,
945, B50, B59, B70, B96, C36, C86, H45, I57, I72, I88, and J01,
which cover this time span or parts of it, show that, within the
errors of measurement, the comet's nuclear brightness was during
the two weeks either nearly steady or somewhat subsiding.

The only other instances of brightening detected in the nuclear
magnitudes from more than one observing site in Table 5 are found
in mid-November:\ between November 13 and 15 from site H47, between
November 14 and 17 from site 958, and between November 17 and 18
from site H45.  Nominally, this looks like a pair of events:\ the
constraint from site 958 is consistent with that from site H47
{\it or\/} H45, but the constraints from H47 {\it and\/} H45 do
not refer to the same event.  Again, no brightening in this general
range of time is apparent in the data from sites 213, 945, B59, C23,
C86, G26, I57, I72, I79, I88, J01, and J38.  Only isolated instances
of brightening are suggested by the data from single sites:\ between
October 10 and 11 from site 850, between October 15 and 16, between
November 11 and 12, and between December 7 and 11 from site 958, and,
finally, between December 3 and 7 from site C86.

Because the second of the two required conditions near the end of
Sec.\ 2.2 is not satisfied, the above account of the suspected cases
of brief brightening in Table 5 provides no evidence on outbursts
after October 1.  These instances could perhaps be explained either
as very brief minor fluctuations of near-nucleus activity or as due to
instrumental/data-reduction problems, including a possible interference
by a field star or stars, whose contribution was not properly removed
from the measured signal.  The broad event between late October and
early November likewise cannot be an outburst because of the enormous
incompatibility of the data from different observing sites.  Its true
nature cannot readily be established from mere inspection of Table 5,
and a different approach is implemented below.  Toward that end,
I next comment on the factors that determine the measured nuclear
magnitudes published by the MPC and then assess the usefulness of
these data beyond their initially recognized role as discriminators
in the applied method for determining the outbursts' onset time.
\begin{table*}
 \centerline{
 \scalebox{0.802}{
 \includegraphics{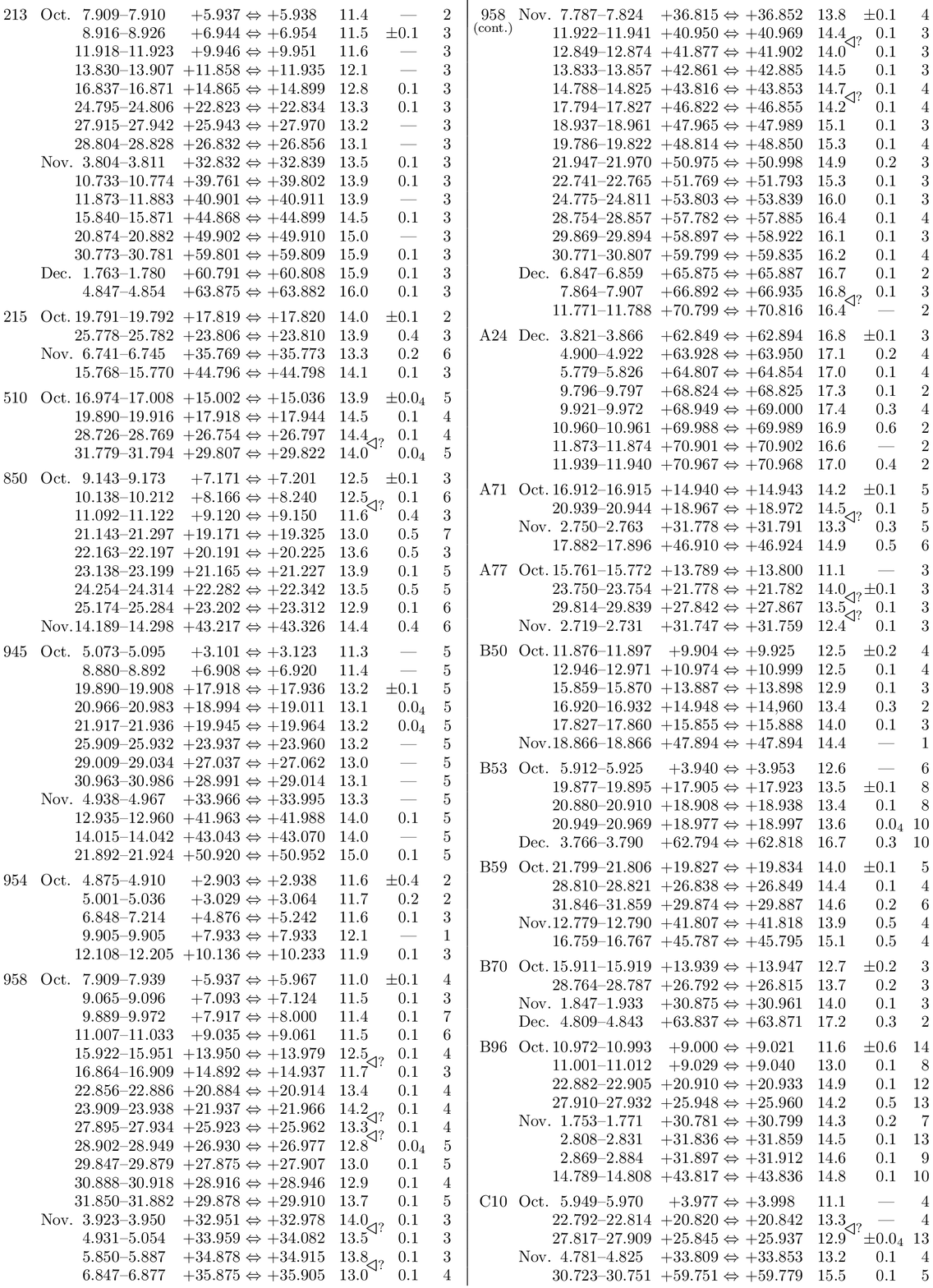}}} 
 \vspace{0.1cm}
\end{table*}
\begin{table*}
 \centerline{
 \scalebox{0.802}{
 \includegraphics{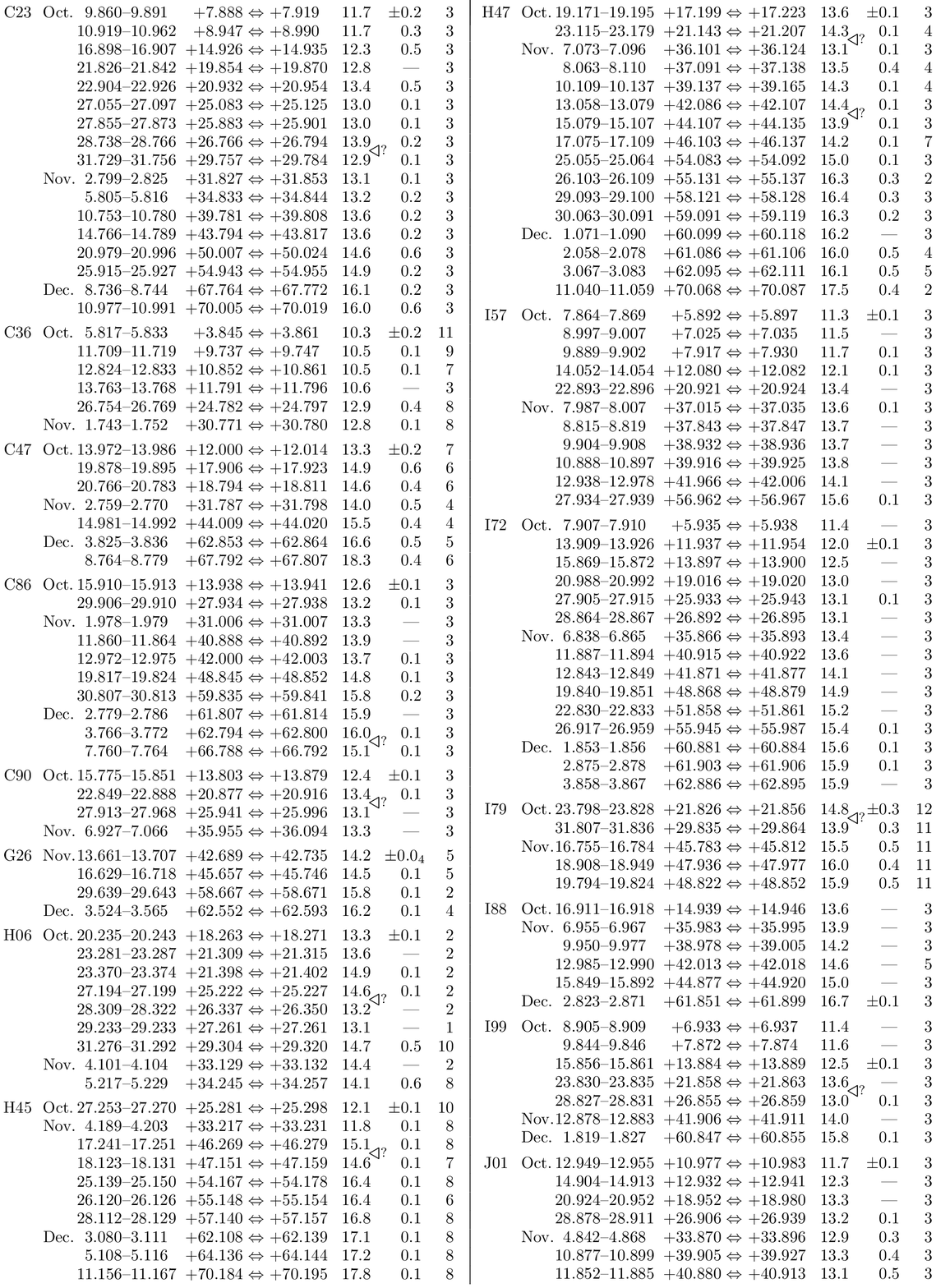}}} 
 \vspace{0.1cm}
\end{table*}
\begin{table*}
 \centerline{
 \scalebox{0.802}{
 \includegraphics{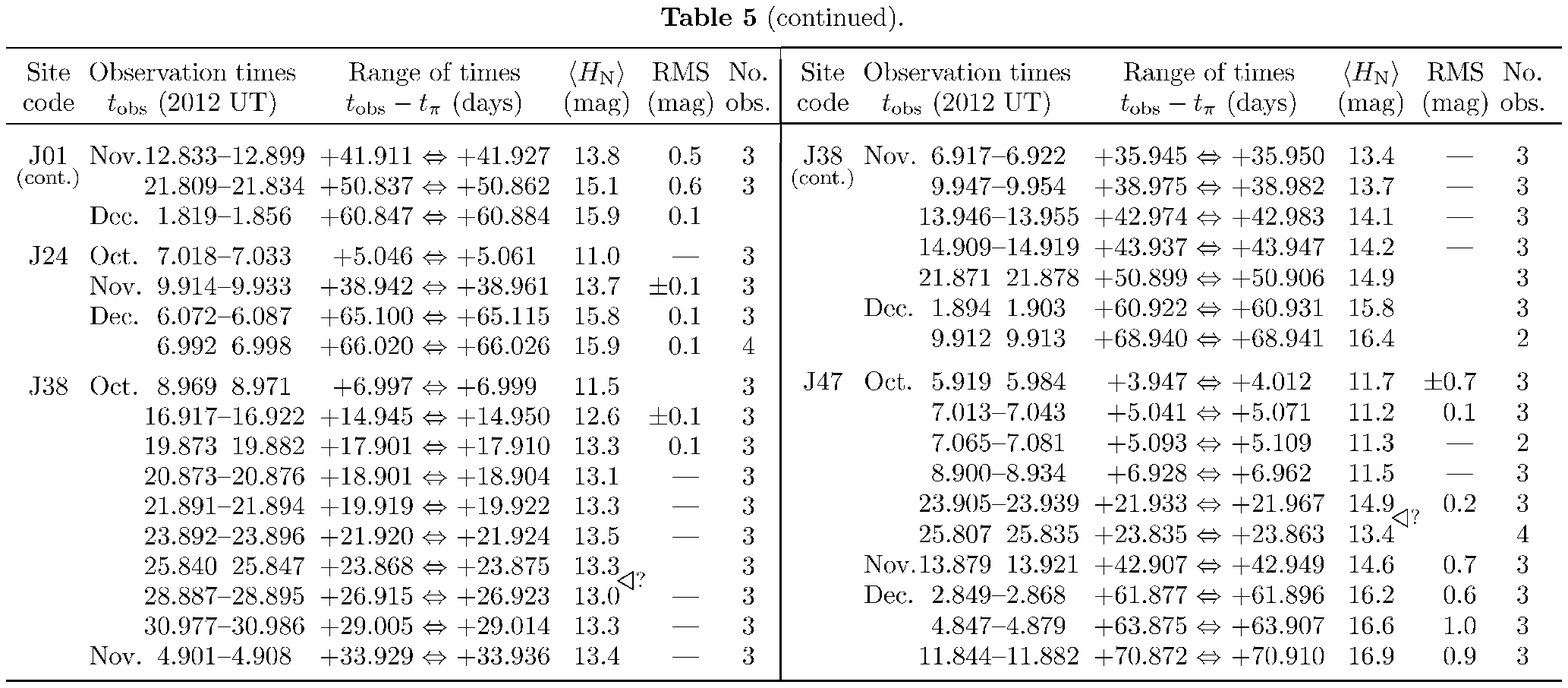}}} 
 \vspace{-9.9cm}
\end{table*}

\begin{table*}
 \hspace{-0.7cm}
 \centerline{
 \scalebox{0.805}{
 \includegraphics{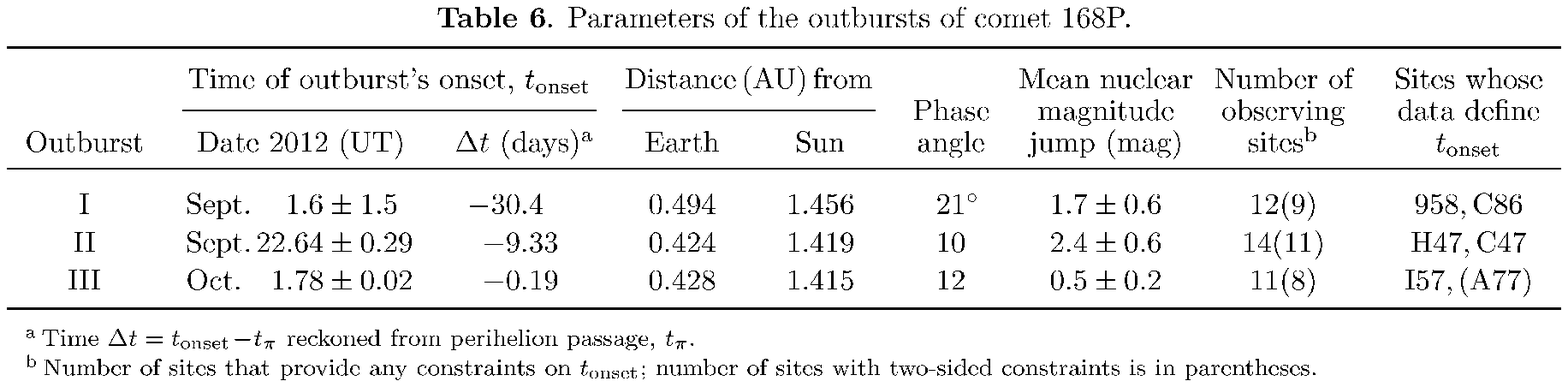}}} 
 \vspace{-9.65cm}
\end{table*}

\subsection{More Information from the Nuclear Magnitudes}

The general feeling of perplexity surrounding the physical
meaning and use of the nuclear magnitudes of comets presented
in the {\it Minor Planet Circulars\/} and the {\it Minor Planet
Electronic Circulars\/} (Sec.\ 2.2) stems primarily from the
uncertainty as to what volume of the inner coma do they refer
to.  The nuclear magnitude of a comet's inner coma (or nuclear
condensation) measured within a circular aperture centered in
a CCD image on the nucleus describes an amount of radiation
coming from a cylindrical volume of space whose diameter at
the nucleus depends --- besides the technical characteristics
of the CCD sensor ---  on:\ (1) the comet's geocentric distance,
(2) the focal distance of the telescope used, (3) the
wavelength-dependent sensitivity of the telescope setup (color
filter used with the CCD chip), (4) the pixel scale, and (5)
the chosen pixel size of the sampling aperture by the person
who reduces the imaging data.

Unfortunately, the format of the MPC astrometric reports of
comets does not provide information included in points (1)
and (3) through (5).  While the geocentric distance can
readily be computed from an ephemeris, the facts in the
other three points cannot be recovered and are lost.  Even
worse, for the observing sites with multiple instrumentation
the report format fails to indicate which observations were
made with which telescope.

There are only two pieces of information that can be invoked
to get at least a crude idea on the volume of space sampled
by the nuclear magnitudes.  One, in the {\it Guide to Minor
Body Astrometry\/}\footnote{See the information web site of
the {\it IAU Minor Planet Center\/} {\tt
http://www.minorplanetcenter.net/iau/info/Astrometry.html.} A
detailed description of the issues related to CCD astrometry~and
photometry of comets is given in Green (1997a, 1997b).} it is
recommended that the pixel scale not exceed, preferably,
2$^{\prime\prime}$/pixel or, at worst, 3$^{\prime\prime}$/pixel,
while simultaneously maintaining a high enough signal-to-noise
ratio.  And, two, in an attempt to standardize the procedure at
least to some extent, the use of an aperture 10$^{\prime\prime}$
in radius was proposed by Kidger (2002).  As long as these two
rules are followed, one finds that the inner coma of up to about
3300 km from the nucleus in the direction perpendicular to the
line of sight contributed to the nuclear magnitude, when comet
168P was at geocentric distances near 0.45 AU (an average of the
geocentric distances at the onset times of the three outbursts;
cf.\ Table 6) and that the diameter of this field should be
covered by 7 to 10 pixels.  The median imaging scale of the
telescopes listed in Table 1 is about 150$^{\prime\prime}$/mm,
so that the preferable pixel scale is satisfied, on the average,
with a pixel size of approximately 13--14 microns on a side,
comparable to that of commonly available CCD arrays.  However,
a few instruments in Table 1 have imaging scales more than twice
as large as the median, and for these even the worst acceptable
pixel scale, 3$^{\prime\prime}$/pixel, requires CCD arrays with
pixels smaller than 10 microns on a side.

Assuming conservatively that the plume of ejecta from the nucleus
of comet 168P expanded at a rate of a few hundred meters per second,
a very brief burst of material (unconsequential to the physical
conditions at the source) released in a direction perpendicular to
the line of sight should have passed through a 10$^{\prime\prime}\!$
aperture in a matter of several hours at the most.  Even if the
direction of the plume's motion was fairly close to the line of
sight, the material should have been out of the 10$^{\prime\prime}$
aperture within one or two days, and the nuclear magnitude should
then have returned to the pre-outburst level.  However, if the
emission event was not brief, the plume of material would have
stayed within the limits of the sampling aperture longer, depending
upon the event's duration.
\begin{figure*}
\hspace*{-0.3cm}
 \centerline{
 \scalebox{0.68}{
 \includegraphics{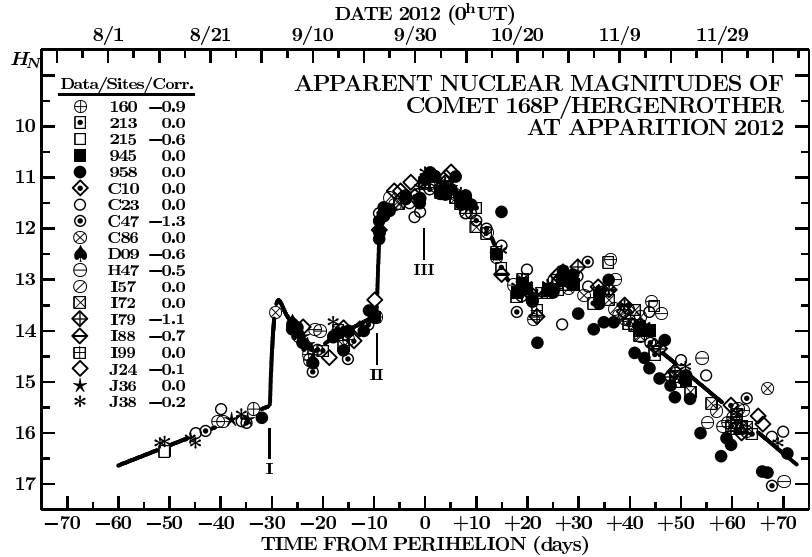}}} 
\caption{Temporal variations in the apparent nuclear magnitude of
comet 168P/Hergenrother, derived from CCD observations obtained
between August 11 and December 11, 2012, or 52 days before perihelion
and 70 days after perihelion, and based on the reports from 19 sites.
The data are referred to the magnitude system used at observing site
958; those from nine other sites required no corrections, while those
from nine other sites were fainter and were corrected by 0.1 to 1.3
magnitudes to become comparable with the rest.  The total number of
plotted points is 303.  The onset times of the three outbursts are
identified by the roman numerals.  A growing scatter among the nuclear
magnitudes from different observing sites is noted toward the end of
the period.\\[0.2cm]}
\end{figure*}

To estimate the strength of the three outbursts, one needs to
study temporal variations in the nuclear magnitudes listed in
Tables 2--5.  I began with site 958, which provided the most
extensive data set.  Abiding by the rule in Sec.\ 2.2 that
nuclear magnitudes from different sites should not be mixed without
first carefully testing them for compatibility, I compared each
of the available nuclear-magnitude sets against the set from
site 958, and was able to distinguish three groups of data:\
(A)~from the sites whose nominal nuclear magnitudes turned out
to be fairly consistent with those from site 958 over the entire
time span, August 11--December 11, 2012, but especially before
October 20; (B)~from the sites whose nominal nuclear magnitudes
could be made fairly consistent with those from group A during
the whole time span after a constant correction has been applied
to the reported nuclear magnitudes; and (C)~from the sites whose
nominal nuclear magnitudes could not be made consistent with the
data from groups A and B without time dependent corrections.  The
classification is not absolute in that especially sites with large
sets of observations, most (but not necessarily all) of which
satisfied the rules for group A or B, were included in that group.
Next to site 958, the sites in group A are 213, 945, C10, C23,
C86, C90, I57, I72, I99, and J36; those in group B are 160, 215,
C47, A24, D09, G26, H47, I79, I88, J24, and J38; and those in
group C are 510, 585, 850, 954, A71, A77, B50, B53, B59, B70, B96,
C36, H06, H45, I81, J01, J08, and J47, some of which offer the
magnitudes only from October or November on (Table~5).  The totals
are 11 sites in group A, 11 sites in group B, and 18 sites in
group C.  The nuclear magnitudes from most group~A and group B
sites are plotted against time in Figure 2, including all such
magnitudes from Tables 2--4.  Only the magnitudes from three such
sites in Table 5, spanning short time periods, are omitted from
Figure 2 (site C90 of group A and sites A24 and G26 of group B).
It is clear that the restrictions on the sets of nuclear magnitudes
that could be incorporated into their common light curve, while
not very tight, prevent the data taken at nearly one half of all
sites from being employed in Figure 2.

This figure allows one to make a number of fundamental conclusions
about the near-nucleus activity of comet 168P.  Outbursts I and
II are prominently displayed, consistent with the large
nuclear-magnitude jumps listed in Table 6.  Outburst III is by
no means striking, but still detectable.  Figure 2 shows that the
shape of the light curve in the aftermath of each of the three
outbursts is quite different.  The nuclear brightness is seen to
have dropped rather steeply starting not later than September 5.9 UT,
some 3--6 days after the onset of outburst I, suggesting that this
event was a relatively brief one, with the active stage spanning
hardly more than two days and possibly only a fraction of a day.
However, the nuclear brightness did not return to the low-activity,
pre-outburst phase, but stayed elevated by at least one magnitude
until the onset of outburst II, which occurred three weeks later.
Once this event commenced, practically no fading is detected in
Figure 2 for about 7 days, so outburst II was more extended in time
than outburst I.  After a brief, shallow drop around September 30,
the nuclear brightness began to climb again sharply on October 1,
the onset of outburst III.  Some sites in Table 4 indicate that
this event was relatively brief, less than two days, while others
suggest that the brightness plateau extended over as many as five
days.  On the average, the peak was reached about October 3 and, in
any case, the comet's activity surely began to subside by October 9.

From Figure 2, the approximate amplitude is 1.9 magnitudes for
outburst I, 2.6 magnitudes for outburst II, and not more than 0.8
magnitude for outburst III.  These amplitudes
are clearly correlated with the average magnitude jumps in Table 6,
exceeding them by 0.2 to 0.3 magnitude, but they are not directly
related to the amount and mass of the material ejected in each
event because the magnitude scale is logarithmic.  In arbitrary
brightness units, the estimated amplitudes correspond to the peak
rates of surge in a ratio of 2, 20, and 15, respectively, for the
three events.  Thus, outburst II was the most powerful one in terms
of both the peak brightness surge and the duration.

As for the category of the outbursts, I have been unable to find any
information on changes in the gas-to-dust ratio potentially associated
with outburst I.  Only incomplete data are currently available for comet
168P on temporal variations in the product \mbox{$Af\rho$}, introduced by
A'Hearn et al.\ (1984) as a proxy to measure the abundance of dust in the
coma.  Sostero et al.\ (2012) calculated \mbox{$Af\rho$} to equal 670~cm
on September 26.6 UT, 1210 cm on October 3.6 UT, and 850 cm on October
9.6 UT within about 3000 km of the nucleus on CCD images (plus a red
filter) taken with the 200-cm f/10 Ritchey-Chr\'etien Faulkes-South
reflector at Siding Spring.  An expanded sample of \mbox{$Af\rho$} values
by G.\ Sostero, G.\ \mbox{Milani}, and E.\ Bryssinck, referring to a
circular aperture of 10,000~km in radius at the comet, is available on
the {\it comets-ml\/} website.\footnote{At
{\tt http://tech.groups.yahoo.com/group/comets-ml}, mes\-sage number
20\,108, date November 6, 2012.}  Although this graph contains about
30 data points, only five of them are in the relevant 30-day window
between the beginning of September and early October.  On September 12
and 14, about midway between outbursts I and II, as well as on September
22.0 UT, shortly before the onset of outburst II, \mbox{$Af\rho$} was
merely 50~cm,\footnote{This value is by a factor of 2.5 smaller than a
median among the $\sim$40 short-period comets in A'Hearn et al.'s (1995)
sample, but much larger than the values listed for 2P/Encke, 10P/Tempel,
26P/Grigg-Skjellerup, or 45P/Honda-Mrkos-Pajdu\v{s}\'akov\'a.} increasing
to \mbox{$350 \pm 70$}~cm on September 26, four days after the onset of
outburst II, and to the rather impressive \mbox{$1500 \pm 300$}~cm on
September 28, three days before the onset of outburst III.  Five days later,
on October 3, \mbox{$Af\rho$} dropped back to \mbox{$740 \pm 150$}~cm
and its nominal value in Sostero et al.'s plot, which extends to early
November, never exceeded 700~cm after October 5 and 200~cm after October
20.  Finally, in an account of his photometric observations on October 9,
Schleicher (2012) gives \mbox{$Af\rho \simeq 300$}~cm, about a factor of
two smaller than the value shown by Sostero et al.\ on their plot.  These
numbers can be compared with the comet's spectrum by C.\ Buil, who, also
on October 9, reported a strong continuum, with only a CN band at 3883
{\AA} and faint [O\,{\scriptsize I}] lines at 5577~{\AA} and 6300~{\AA}
being detected.\footnote{Full description of Buil's observation, made
with his 23.5-cm f/10 Celestron C9 Schmidt-Cassegrain reflector, can be
found at {\tt http://www.astrosurf.com/buil/comet/hergenrother/obs.htm}.}

Based on all this evidence, outbursts II and III appear to have been
dust dominated events.  Although no relevant information is available
for outburst I, it is probably rather safe to speculate that it too
was dominated by dust.

Figure 2 also provides information on the shape of the broad event in
late October and early November.  Clearly the fairly rapid drop in the
near-nucleus activity that followed outburst III ceased on or around
October 21 and in the following two or so weeks the nuclear brightness
either stagnated or even surged up a little, suggesting possibly a
limited re-activation of the nucleus.  The data from the various
observing sites do not provide a consistent answer as to what exactly
was happening, but the event certainly was not a major flare-up.  After
November 7 the fairly steep rate of brightness decrease generally resumed,
continuing until the end of the investigated period of time, 70 days
after perihelion.

\section{Fragmentation of Comet 168P/Hergenrother and a Correlation
of the Separation Times with the Onset Times of the Outbursts}
The first report of a secondary nucleus came from Sostero et al.\
(2012) who detected it on stacked images taken with the 200-cm f/10
Ritchey-Chr\'etien Faulkes-North reflector atop Haleakala, Maui, on
October 26.4 UT.  This fragment B was located about 2$^{\prime\prime}$
south and slightly to the west of the primary nuclear condensation
(now fragment A), was of magnitude $\sim$17, and had a diffuse coma
nearly 2$^{\prime\prime}$ in diameter.  Fragment B was still visible
on images taken with the same telescope on November 2 and 3, but not
on November 7, when it must have been fainter than magnitude 20.
However, on this last date Sostero et al.\ suspected another extremely
faint fragment a little more than 8$^{\prime\prime}$ to the southeast
from the primary, which however was not confirmed.\footnote{This
fragment, if genuine, could be either companion C or F; however,
neither fits quite satisfactorily.}

Stevenson et al.\ (2012) reported the results of their observations
of comet 168P with the 810-cm Gemini-North telescope atop Mauna Kea,
Hawaii, on October 28 and November 2.  On both nights they confirmed
the presence of fragment B and on the second night they detected two
additional fragments, C about 6$^{\prime\prime}$ to the southeast of
the primary and D more than 11$^{\prime\prime}$ to the south-southeast
of the primary.

Hergenrother (2012a, 2012b) measured a total of five nuclear companions
in a number of the comet's images taken with the Faulkes-North reflector
on October 26 and November 2--3 and by Y.\ Fernandez and E. Kramer with
the 210-cm telescope atop Kitt Peak between November 2 and 12.  The
details of all measurements of companions B--G employed in the present
investigation are listed in Table 7, which also includes privately
communicated information from Hergenrother on his unpublished measurement
of companion D in the images taken with the 183-cm Vatican Advanced
Technology Telescope atop Mount Graham on November 17.

\begin{table*}
 \hspace{-0.6cm}
 \centerline{
 \scalebox{0.81}{ 
 \includegraphics{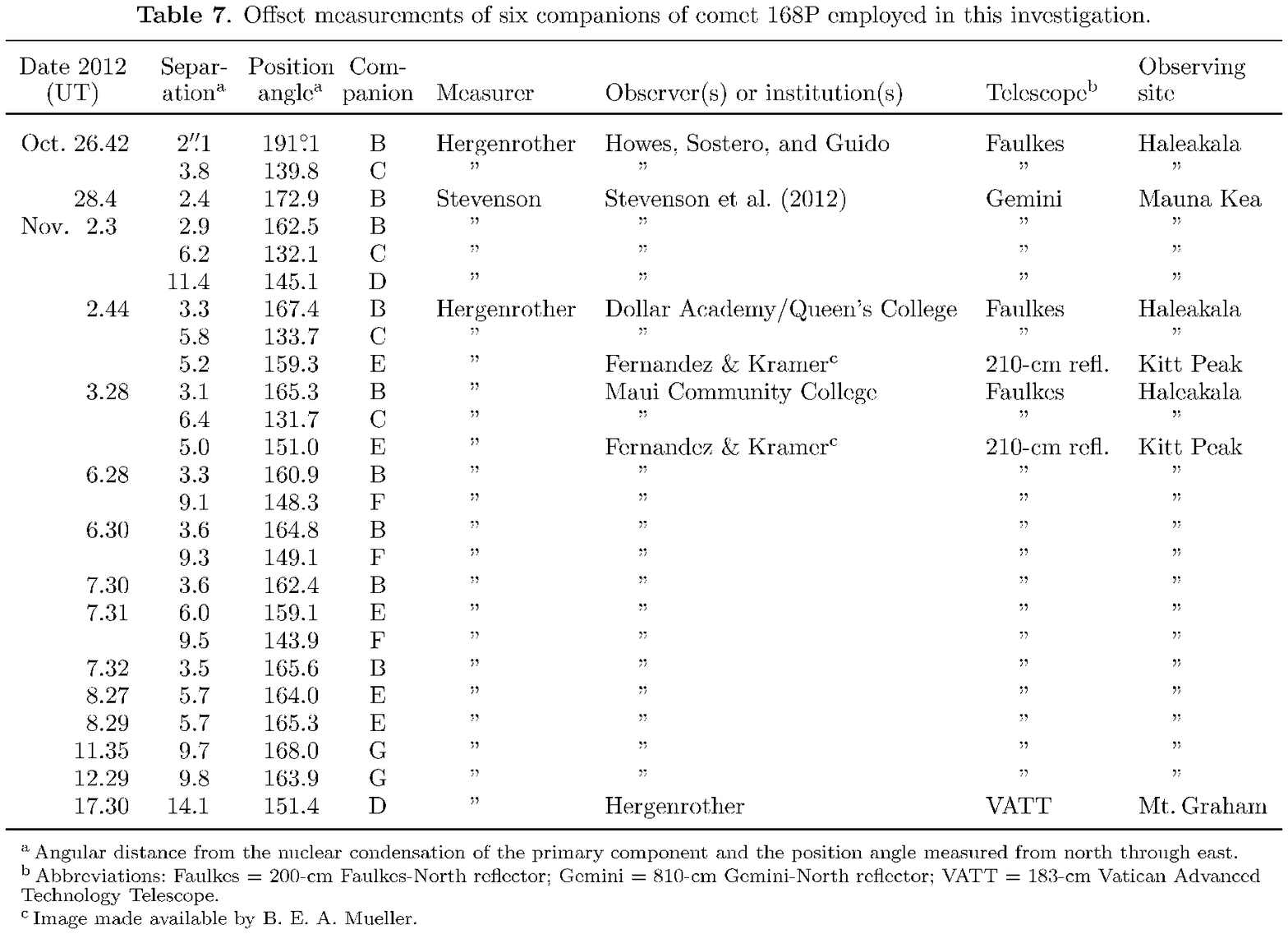}}} 
 \vspace*{-5.95cm}
\end{table*}

\subsection{Determining the Fragmentation Parameters}
The motions of the six companions of comet 168P~are now modeled to
derive the fragmentation parameters, employing the technique developed
by the author more than three decades ago (Sekanina 1977, 1978, 1982)
and extensively tested over the years.  An upgraded version of this
method, which includes the differential perturbations by the planets,
was described by Sekanina and Chodas (2002).

In general, the goal is to determine up to five fragmentation parameters
for a companion separating from the parent comet:\ the time of its
breakup, called the time of separation or fragmentation, $t_{\rm frg}$;
the differential deceleration $\gamma$ due to outgassing; and the
velocity of separation $V_{\rm sep}$ of the companion relative to the
parent at time $t_{\rm frg}$.  The deceleration is assumed to act
continuously in the antisolar direction and to vary as the inverse
square of heliocentric distance.  The right-handed RTN coordinate
system is centered on the parent object, referred to its orbit plane,
and defined by the orthogonal directions radial away from the Sun,
transverse in the plane, and normal to the plane.  The components
of the separation velocity in this coordinate system are $(V_{\rm
sep})_R$ in the radial direction, $(V_{\rm sep})_T$ in the transverse
direction, and $(V_{\rm sep})_N$ in the normal direction.  The
employed iterative differential-correction least-squares optimization
procedure makes use of software that solves the normal equations for
an arbitrary number of unknowns.  The technique thus allows one to
determine all five parameters [\,$t_{\rm frg}$, $\gamma$, $(V_{\rm
sep})_R$, $(V_{\rm sep})_T$, and $(V_{\rm sep})_N$] or any combination
of fewer than five of them; a total of 31 different versions is
available.  This option proves very convenient, especially in an
early phase of the optimization process, before the solution
``settles'' near the most probable values of the parameters, or
when the convergence is slow.  It is also highly beneficial when a
data set is too small to allow one to determine all five parameters.
This feature is in the following calculations used to great advantage.

The primary task for this investigation of the motions of the six
companions is to examine their implied fragmentation times and the
possible correlation between them and the onset times of the three
outbursts.  The number of offset observations in Table 7 is very
limited, so one cannot expect that the full five-parameter model
could be applied, particularly because experience has shown that
the derived radial component of the separation velocity, $(V_{\rm
sep})_R$, is often highly correlated with the fragmentation time
$t_{\rm frg}$.

In an effort to find the best possible fragmentation parameters, I
search for different solutions to the available data set of each
companion.  To mark them apart, I assign each solution a group of
letters, which indicate what parameters are included; F stands
for the fragmentation time, D for the deceleration, and R, T, and N
for the radial, transverse, and normal components of the separation
velocity, respectively.  Thus, for example, solution~FD means that
only the fragmentation time and the deceleration are solved for and
that the separation velocity is forced to be zero; similarly,
solution~DRN means that the deceleration and the radial and normal
components of the separation velocity are solved for, with a particular
forced fragmentation time and a zero transverse component of the
separation velocity.  Various solutions are compared in terms of the
root-mean-square error and whether or not the systematic trends with
time are present in the distribution of residuals ``observed minus
modeled'' in both the right ascension [which includes the factor
cos\,(declination)] and the declination.  These criteria serve to
assist in judging the quality of the solutions and in facilitating
the final choice of the individual parameters, primarily the
fragmentation time.  And because the outgassing-driven deceleration
has a dominant effect on the motion of any companion, the first step
in the process of estiamting the most probeble time of fragmentation
is to compute the solutions that involve these two parameters.

It should also be noted that the earth transited the orbit plane of
comet 168P on September 19.2 UT, 2012, which resulted in unfavorable
edge-on observing conditions in the days around this time, as any mass
released from the nucleus in, or very close to, the orbit plane was
in projection onto the plane of the sky moving in the same direction.
Fortunately, thanks to the comet's 22$^\circ$ inclination and
relatively small geocentric distance, the earth's angular distance
from the plane began to increase fairly rapidly soon after the
transit, reaching 10$^\circ$ on September 30, 20$^\circ$ on October
15, and 25$^\circ$ by the time the first companion was detected,
on October 26.

\subsection{Companion B}
Table 7 shows that nine measurements of the offsets of this fragment
from the primary A in the images taken between October 26 and November 7
are available for computing the fragmentation parameters.  Three
solutions that included the fragmentation time as a variable, FD, FDT,
and FDN, are listed in the first three lines of Table 8.  Solution FDR
and three four-parameter solutions, FDRT, FDRN, and FDTN, failed to
converge.

\begin{table*}
\hspace{-0.45cm}
 \centerline{
 \scalebox{0.797}{ 
 \includegraphics{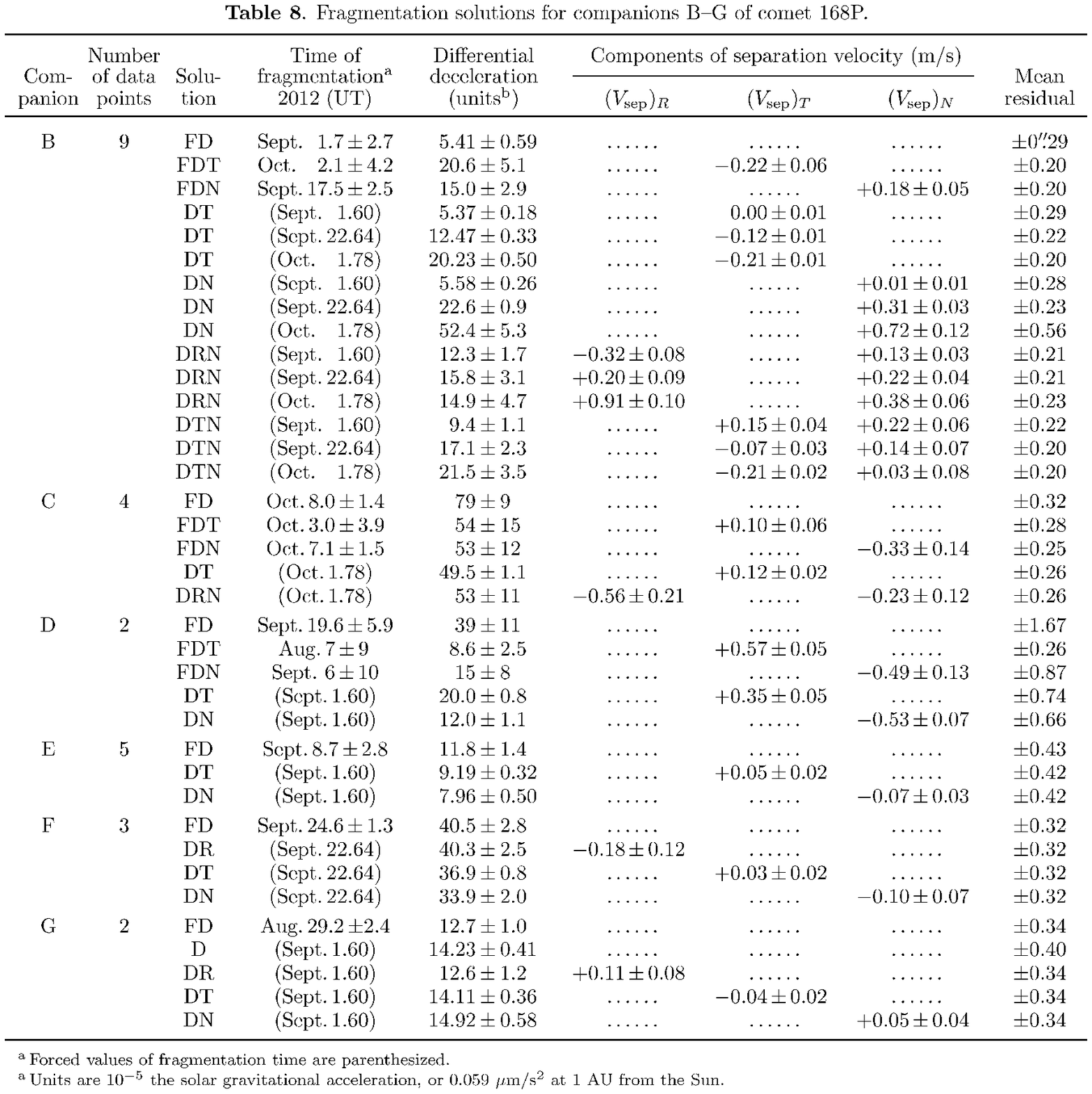}}} 
\vspace{-3.95cm}
\end{table*}

By sheer coincidence, in the three runs, in which the fragmentation
time was solved for, the computed values of this parameter just happen
to span the period covered by the three outbursts, thus providing no
obvious clue as to which of them is the one most probably associated
with the release of this companion.  However, both the comparison of
solutions FD, FDT, and FDN in Table 8 and the distribution of residuals
from solutions FD and FDN in Table 9 (which also shows the offsets
in right ascension and declination), slightly favor a fragmentation
time in the second half of September or in early October, so that
outburst I is a less likely candidate.  Table 9 also presents the
residuals from other solutions of particular interest, based on
three values of the fragmentation time forced to coincide with the
onset time of each of the three outbursts.  Solutions DN and DRN appear
to prefer outburst II, while solution DT favors slightly outburst III,
and solution DTN is essentially inconclusive.  Thus, by an extremely narrow
margin, outburst II may be the most likely one to correlate with companion B.
The DR-type solutions are not listed in Table 8, because they always
resulted in an inferior fit, with the mean residual of about
$\pm$0$^{\prime\prime}\!$.3 or more.

\begin{table*}
\hspace{-0.4cm}
 \centerline{
 \scalebox{0.8}{
 \includegraphics{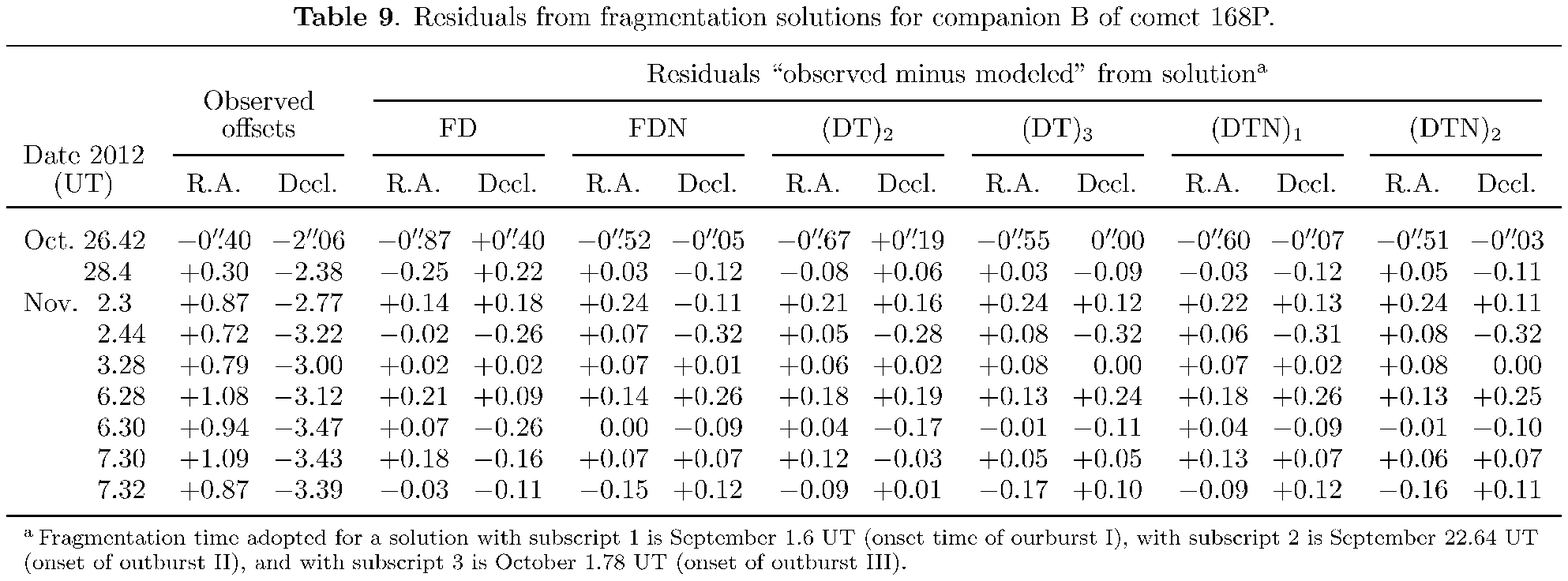}}} 
 \vspace{-8.5cm}
\end{table*}

\subsection{Companion C}
Offsets of this companion from the primary were measured only in the
images from four days, between October 26 and November 3 (Table 7).
The three solutions, including the fragmentation time as a variable,
provide for it the dates of October 3--8 with a 1$\sigma$ uncertainty
of up to nearly 4 days, as shown in Table 8.  The likely correlation
with outburst III is supported by the two solutions, DT and DRN, that
were run on the assumption that the release of companion C coincided
with the onset of outburst III.  Somewhat surprisingly, solution DTN
had a somewhat larger mean residual than solution DRN and is not
listed in Table 8.  Judging from their mean residuals, solutions DRN
and DT are both acceptable.  Solution DT also offers a reasonably
good distribution of residuals in Table 10.  One can conclude with
some confidence that a close relationship between companion C and
outburst III is quite plausible.  

\subsection{Companion D}
This fragment was extremely faint on both November 2 and 17 and,
curiously, was not detected in between the two dates.  With only
these two data points, one is extremely limited in terms of choice
of solutions.  The large gap between them also offers an opportunity
for contradictory solutions that may provide an unexpectedly good fit
to the two points but lead to fictitious fragmentation parameters
and must be rejected.

Table 8 presents the three standard types of solutions that include
the fragmentation time.  The simplest, solution FD, leads to the
separation of D on September 19, fails to fit the November 2 offset
in right ascension by fully 2$^{\prime\prime}$, and is therefore
unacceptable.  Solution FDT, although by far the best of the five
in Table 8 in terms of fitting the two data points. is also
unacceptable, because it implies the fragmentation time long before
the activation of the nucleus.  The FDN solution points to early
September as the most likely fragmentation time, which is confirmed
by even better fits based on solutions DT (Table 10) and DN, in which
the coincidence was assumed between the fragmentation time and the
onset time of outburst I.  This outburst is likely to have accompanied
the birth of companion D.

\begin{table}[b]
\vspace{0.5cm}
\hspace{-0.41cm}
 \centerline{
 \scalebox{0.78}{
 \includegraphics{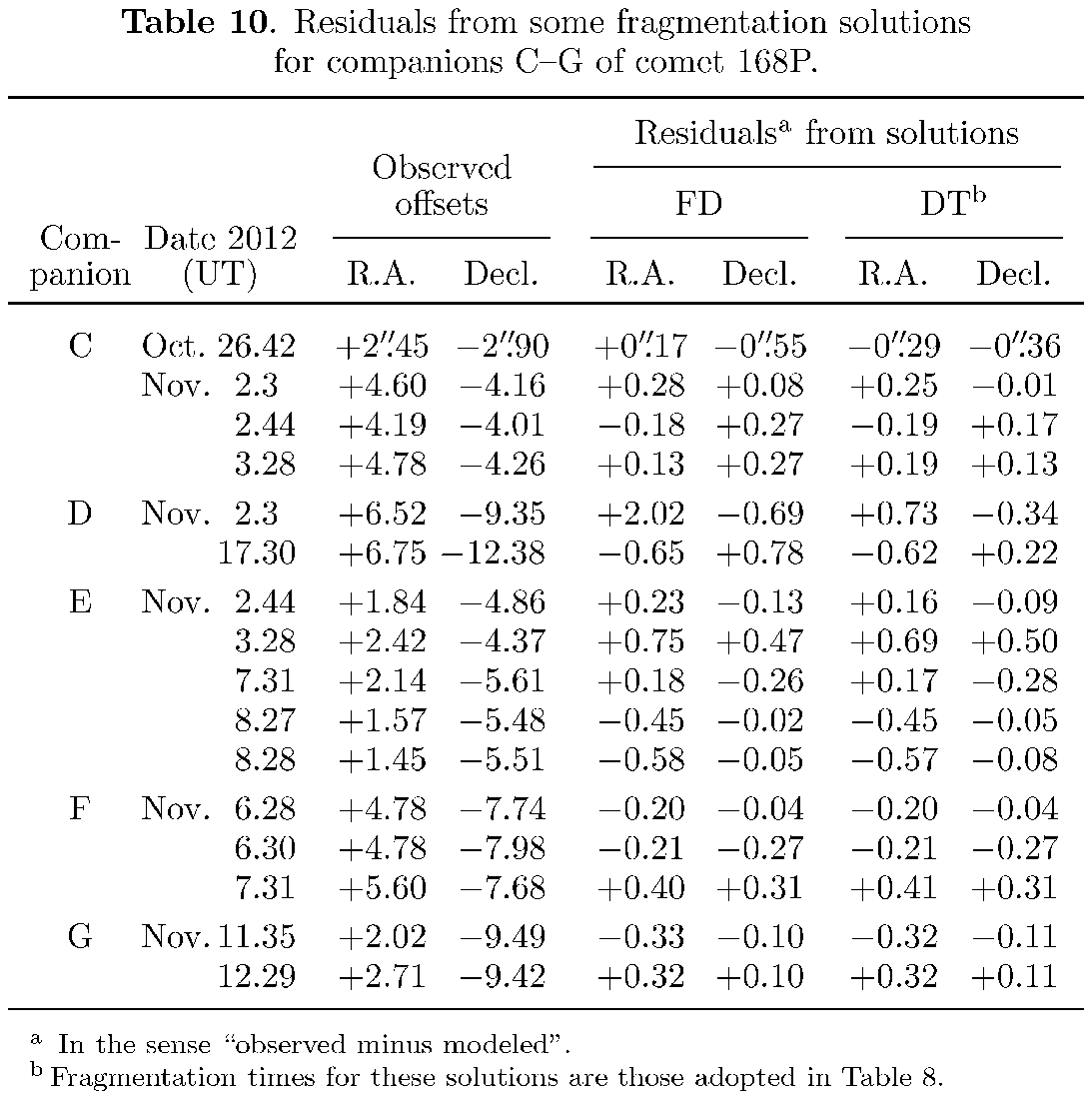}}}
\vspace{-7.2cm} 
\end{table}

\subsection{Companion E}
Even though this companion was measured on four nights, it was possible
to derive the fragmentation time only from solution FD (Table 8), but
not from FDT or FDN, both of which failed to converge.  This problem was
the reason for abandoning, in the prelimnary report (Sekanina 2012a),
the scenario of companion E having been released from the primary and,
instead, preferring it to be a fragment of companion B.  With the
present, much more extensive investigation, the offsets of E from what
is now considered the best possible solution for B provide less
attractive solutions to such a fragmentation scenario.  While no
solution for E relative to the primary offers an entirely satisfactory
distribution of residuals, one has no choice but to accept as adequate
solutions DT or DN, based on the assumption of this companion having
been released from the primary at the onset time of outburst I (Table 8).
Both solutions yield rather similar residuals; the ones for solution DT
are listed in Table 10.  Identifying the fragmentation time with the
onset time of outburst II or III leads to substantially inferior solutions,
with the mean residuals near $\pm$0$^{\prime\prime}\!$.5 or worse and with
strongly systematic distributions of residuals.

\subsection{Companion F}
By contrast, the three measurements of this companion on two consecutive
nights left little room for a broad variety of fragmentation scenarios.
Solution FD, the only possible one that includes the fraagmentation time
as a variable, suggests rather unambiguously that this fragmentation
event was related to outburst II (Table 8).  Indeed, the table also
shows three equivalent two-parameter solutions, DR, DT, and DN, which
were obtained by forcing the fragmentation time to coincide with the
onset time of this outburst.  The residuals, acceptable under the
circumstances, are in Table 10.  No three- or four-parameter solutions
could be made to converge.

\subsection{Companion G}
Only two images of this companion were measured on two consecutive nights,
and the choice of fragmentation scenarios was as limited as in the case
of companion F.  Solution FD in Table 8 suggests that the birth of this
companion was related to outburst I.  When only the deceleration was
solved for, the fit deteriorated a little, but equivalent two-parametric
solutions in which the fragmentation time was forced to coincide with
the onset time of outburst I had the same mean residual and the
individual residuals virtially identical with those from solution FD
(Table 10).

\begin{table}
\vspace{0.15cm}
\hspace{-1.85cm}
 \centerline{
 \scalebox{0.8}{
 \includegraphics{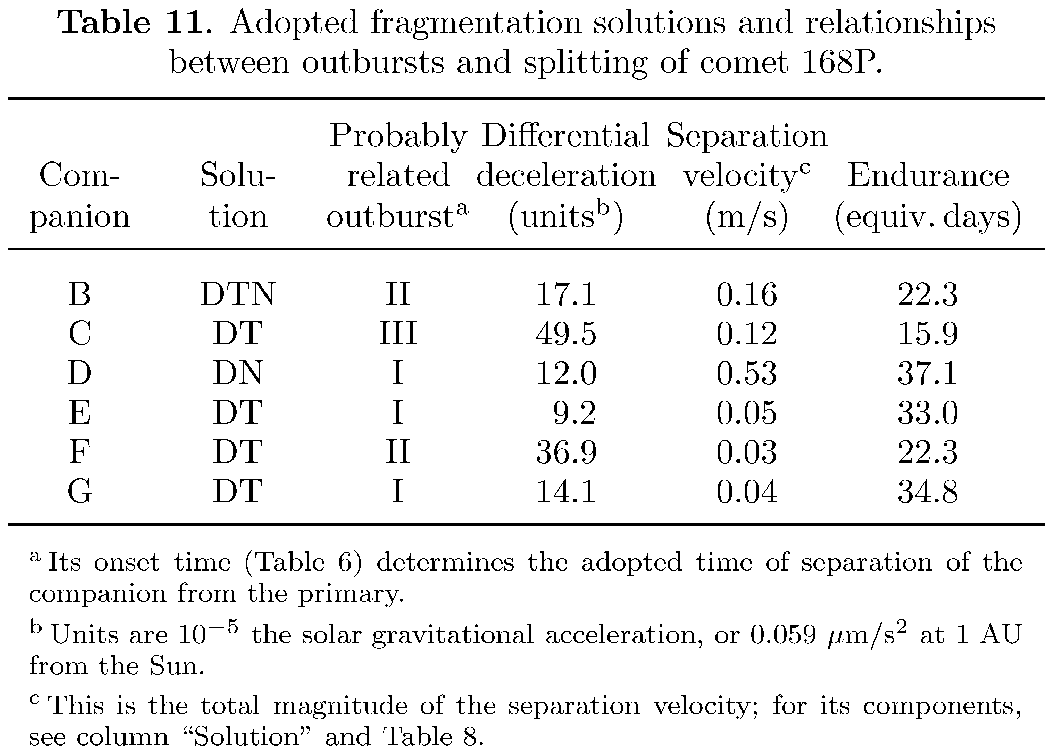}}} 
\vspace{-8.5cm}
\end{table}

\subsection{Summary of Findings on the Fragmentation Events}
It is unfortunate that all six detected companions of this comet have
been short-lived, none surviving for more than 11 weeks and mostly much
less than that.  The rather irritating experience with their appearance
confirms that they all were typical cometary fragments in that they
underwent dramatic brightness fluctuations with time, having been
brighter than about magnitude 20 on only rather rare occasions.  As
a result, their observations were very scarce and their positional
measurements exceedingly difficult.

It is highly probable that all six companions separated directly from
the primary.  The fragmentation solutions offered in Table 8 show,
however, that because of the scarcity of observations it was never
possible to solve for all five parameters or even four parameters of
the fragmentation model.  One of the corollaries of this problem is
a greater than expected error in the fragmentation times derived.
This is true for all six fragments, including B, the best observed
one.  Under these circumstances, one cannot expect to {\it prove\/}
that the times of the fragmentation events truly coincided with the
onset times of the outbursts.  Rather, one needs to take it for granted
that for comet 168P the two categories of phenomena {\it were\/} closely
related and, on this assumption, try to figure out which outburst
might have accompanied each of the fragmentation episodes.

This objective was for each companion discussed in the preceding subsections.
The adopted fragmentation solutions and the relationships between the
outbursts and the fragmentation events are summarized in Table 11.  The
most likely scenario that emerges from this table is that outburst I
coincided with the separation of three companions, while outburst II
accompaned the birth of two companions and outburst III just one companion.
The relationships between outburst I and companion G, between outburst II
and companion F, and between outburst III and companion C are proposed
with somewhat greater confidence than the relationships between outburst I
and companions D and E, and between outburst II and companion B.

These assignments, if correct, are remarkable in that the least powerful
explosion event of the three, outburst~I (Sec.\ 3.2), correlates with
three companions, while the most powerful one, outburst II, with only two
companions.  Thus, the amount of ejected dust appears to be inversely
correlated with the mass released intact at least when comparing
outbursts I against II and I against III.  This could mean that the
total mass of the lost solid material may not have varied dramatically
from event to event, but its overall mechanical strength may have.

The last column of Table 11 presents the observed endurance $E$ of
the companions, defined (Sekanina 1977, 1982) as the interval of
time, $t_{\rm frg} - t_{\rm last}$, between fragmentation and the
last observation, weighted by an inverse square power law of
heliocentric distance $r$, thereby measuring each fragment's
minimum lifetime against its outgassing, whose rate is approximated
by the $r^{-2}$ law.  Thus,
\begin{equation}
E = \int_{t_{\rm frg}}^{t_{\rm last}} \! \frac{dt}{r^2} = 1.015 \,
   p^{-\frac{1}{2}} \! \left( u_{\rm last} - u_{\rm frg} \right),
\end{equation}
where $p$ is the semilatus rectum of the fragment's orbit (which for
all fragments of 168P can be approximated by the value of $p$ of the
comet's orbit) and $u_{\rm last}$ and $u_{\rm frg}$ are the true
anomalies at the times of, respectively, the last observation and
fragmentation.  When $p$ is in AU and the true anomalies in degrees,
the endurance is expressed in equivalent days, that is, the days at
1 AU from the Sun.

The plot of the endurance $E$ of 24 companions of 18 split comets
against their differential deceleration $\gamma$ shows (Sekanina
1982) that $E$ generally increases with decreasing $\gamma$ and that
this relationship is described by
\begin{equation}
E = \Lambda \gamma^{-0.4},
\end{equation}
where $\Lambda$ is a constant and $\gamma$ is in units of 10$^{-5}$\,the
solar gravitational acceleration.  A great majority of fragments of the
split comets follows this empirical law with \mbox{$\Lambda = 200$ equivalent}
days.  However, a small group of sturdier, relatively massive objects
\mbox{($\gamma < 10$ units)} satisfies law (6) with \mbox{$\Lambda \simeq
800$ equivalent} days, while another group of five very brittle, low-mass
fragments, whose \mbox{$\gamma > 60$ units}, fits law (6) with \mbox{$\Lambda
= 87$ equivalent} days.  The surprising finding from Table 11 is that all
six companions of comet 168P match closely an extrapolated $E(\gamma)$
relationship of this group of very brittle fragments, as seen from
Figure 3.

The five fragments in this group that broke off from earlier comets are,
in the order of decreasing $\gamma$, companion C to C/1899~E1 (Swift) and
companions B to C/1906~E1 (Kopff), C/1942~X1 (Whipple-Fedtke-Tevzadze),
C/1968~U1 (Wild), and C/1969~T1 (Tago-Sato-Kosaka).  The first two are
Oort-cloud comets, while C/1969~T1 and C/1942~X1 have the original orbital
periods of about 90,000 years and 1600 years, respectively.  The orbit of
comet C/1968~U1 has not been determined to adequate accuracy to establish
its origin.  In any case, 168P is the first short-period comet whose
companions belong to this group of excessively brittle fragments.
\begin{figure}[b]
\vspace{0.4cm}
\hspace{-1.05cm}
 \centerline{
 \scalebox{0.67}{
 \includegraphics{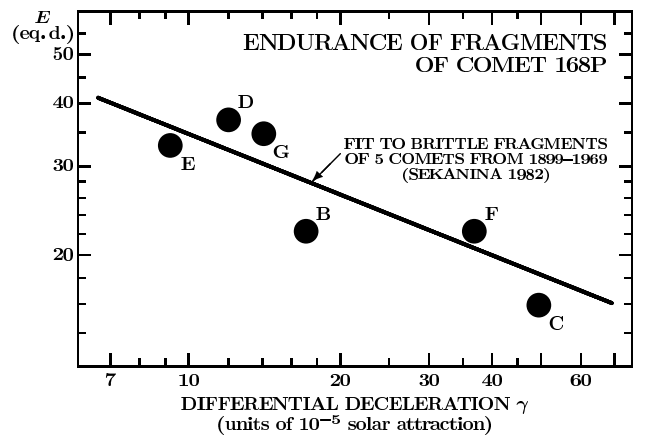}}} 
\vspace{-9.15cm}
\caption{Endurance of the companions of comet 168P as a function of
their differential deceleration due to outgassing.  The straight line
is not a fit to the plotted data points, but extrapolated from that to
very brittle fragments of five comets from 1899--1969.}
\end{figure}

One may further add that this excessive brittleness explains not only the
unusually short lifetimes of the companions of 168P, but it also fits
the observers' reports that the companions displayed generally a tendency
toward progressive elongation, a sign that they already consisted of
expanding clusters of subfragments subjected to a range of decelerations.
This accelerating disintegration of subfragments into boulders, pebbles,
and coarse dust is a characteristic property of the advanced phase of
the process of cascading fragmentation and has recently been under
different circumstances demonstrated by comet C/2011~W3 only days after
its passage through the Sun's inner corona (Sekanina and Chodas 2012).

Thanks to comet 168P, the global picture of the plot of the endurance
$E$ versus the deceleration $\gamma$ changes in the sense that the
previous range of decelerations in this group of excessively brittle
fragments, between 65 and 480~units of 10$^{-5}$\,the solar
gravitational acceleration, is now greatly extended beyond the lower
limit of $\gamma$ all the way to slightly less than 10~units, that is,
the range in $\log \gamma$ has nearly doubled, showing that this group
is by no means limited to merely dwarf fragments.

It is noticed that some results in my first preliminary report on the
fragmentation sequence (Sekanina 2012a) differed from these final
conclusions.  The differences are due in part to the data point on the
companion D until recently unavailable, in part to a more comprehensive
analysis now undertaken.  The results of the second~pre\-limi\-nary
report (Sekanina 2012b) have now been closely confirmed.\\[-0.5cm]

\section{Mass of Material Trailing\\the Nucleus of 168P}
Hergenrother (2012a) called attention to a mass of material appearing,
in numerous images taken in the second half of October, to move away
from the near-nuclear region in the antisolar direction.  First
detected in the high-resolution images exposed with the Faulkes-North
200-cm reflector on October 16, the feature was present until at least
October 23, but the Faulkes images from October 26 no longer show it.
On the very likely premise --- supported by the fairly high values
of $Af\rho$ in this period of time (Sec.\ 3.2) --- that this mass
consisted of dust ejecta, its position angles measured by Hergenrother
are compared in Table 12 with the position angles computed for the
best-fitting synchronic formation.  The corresponding most probable
time of the emitted material is October 5.8\,$\pm$\,0.6 UT.  However,
it was pointed out by Sostero et al.\ in their blog\footnote{The
blog, dated October 22, 2012, by G.\ Sostero, N.\ Howes, A.\ Tripp,
and E. Guido, {\tt http://remanzacco.blogspot.it/2012/
10/update-on-comet-168phergenrother.html}.} that the images acquired
with the Faulkes-North telescope on October 22.44 UT showed this
``diffuse trail'' to be about 6$^{\prime\prime}$ long and
3$^{\prime\prime}$ wide, the width suggesting that the duration of
the emission event was nontrivial, extending perhaps over a period
of a few days or so.  The overall timing of this feature's emission
appears unquestionably to be related to outburst III and the release
of companion C.  The positional correlation between the mass of
trailing material and companion C was noticed by Hergenrother (2012a),
and the suspicion that this mass was a product of outburst III was
expressed by the author in the same communication note (Sekanina
2012a).
\begin{table}[b]
\vspace{0.4cm}
\hspace{-1.95cm}
 \centerline{
 \scalebox{0.8}{
 \includegraphics{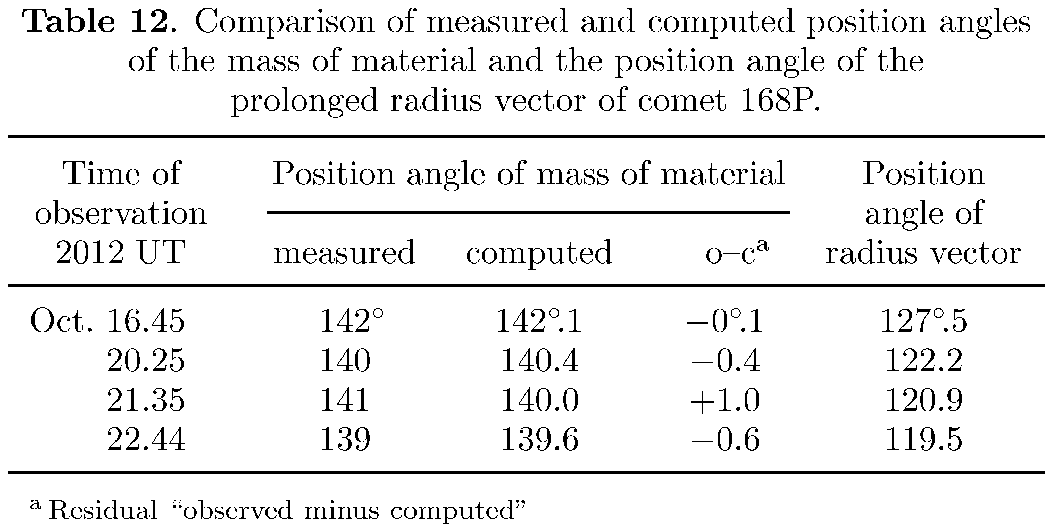}}} 
\vspace{-9.8cm}
\end{table}

The dimensions of dust particles in this trailing mass can be estimated
from the length of the synchronic feature.  Its just mentioned extension
on October 22.44 suggests that the dust was subjected to a maximum
solar radiation-pressure acceleration of $\sim$0.0018 the
Sun's gravitational acceleration, which at an assumed bulk density
of 0.4~g/cm$^3$ is equivalent to a minimum particle diameter of
$\sim$1.6~mm.  Thus, the observed mass was made up of dust mostly
in the millimeter-centimeter size range.  Curiously and perhaps not
quite coincidentally, this limiting particle diameter is identical
to that derived for the dust population situated at the tip of the
spine tail of comet C/2011~W3 after it lost its nuclear condensation
(Sekanina and Chodas 2012).
%

\section{Conclusions}
Comets are notorious for always changing their brightness, but not
every brightening is called an outburst.  To belong to this category
of events, a brightness surge must satisfy three critical conditions:\
it must be sudden, sufficiently prominent (amplitude of at least
0.8--1.0~magnitude), and unexpected.  Countless studies have shown that
the total amount of material ejected from the nucleus during an outburst
is always less (usually far less) than 10$^{13}$\,grams and that an
outburst produces a {\it local\/} event on the scale of the dimensions
of an average comet nucleus.  Outbursts differ from the very rare giant
explosions, which are much more massive and powerful.  During the
outbursts, to which some comets are more prone than other, both dust
and gas are released from the nucleus, but there is a wide range of
these events in terms of the dust-to-gas mass ratio.  It is therefore
appropriate to talk about dust-dominated and gas-dominated outbursts.
Their differences are revealed not only spectroscopically, but also by
their unequal temporal profiles, as the history of their observations
clearly documents.

Because a major attribute of outbursts is a steep light surge in the
initial stage of a ``stellar nucleus'' --- an unresolved image of an
expanding plume of ejected material --- a fundamental parameter of
these events is their onset time.  An extra dimension to this parameter
is provided by the fact that outbursts often coincide with the comets'
fragmentation events, and the onset time can be used to correlate the
two classes of phenomena.  In the absence of accurate information on
the timing of a fragmentation event, an outburst's onset time can
even be used as a proxy for the separation time of a nuclear fragment.
In the case of multiple outbursts and multiple fragmentation of
a comet, this approach can be applied to test various
fragmentation/outburst sequences and scenarios.  This is the case
of comet 168P/Hergenrother.

Next I show that the so-called nuclear magnitudes, published in the MPCs
and the MPECs for comets that are observed by means of CCD arrays primarily
for astrometric purposes, can despite their poor reputation be used to
great advantage in an effort to tightly constrain the onset time of
outbursts in objects that are extensively monitored.  Although it is
inadmissible to mix the nuclear magnitudes reported for the same comet
by different observers without first carefully examining their possible
compatibility, it is legitimate to combine the temporal constraints on
an outburst's onset time derived from timing of images reported by the
various observers, as long as the conditions derived from this timing
independently and consistently confirm the overall outcome and, if so
happens, any minor inconsistencies are fully understood in terms of the
observational setups.

Extensive application of the developed technique to comet 168P shows
that the object underwent three separate outbursts during a one-month
period between the beginning of September and the beginning of October
2012.  Toward the end of October, yet another modest surge of activity
occurred, but it was neither sudden nor prominent enough to be classified
as an outburst.  Afterwards, the comet's activity was steadily decreasing
with no flare-ups worth mentioning, the monitoring having been terminated
before mid-December, 70 days post peri\-helion.  The amplitudes of the
three outbursts were used to estimate their peak rates of brightness surge
in arbitrary intensity units; their ratio was found to be 2:20:15.

High-resolution imaging of comet 168P revealed the existence of six
companion nuclei, B--G, to the primary A between October 26 and
November 17.  The modeling of their motions suggested that they all
broke off from the primary and that the comet was indeed fragmenting
profusely during the period of time covered by the three outbursts.
Whereas the exact fragmentation times could not be established from
the limited astrometric data, closer examination suggested that the
first outburst was most likely to accompany the release of companions
D, E, and G; the second outburst to be associated with the birth of B
and F; and the last outburst to coincide with the separation of C.  The
peak rates of brightness surge do not at all appear to be correlated with
the number of fragments released.  This tendency to an anticorrelation
between fragmentation and the magnitude of outbursts could mean that
the material losses during the three outbursts were comparable in mass,
but that most of the mass separated in the first outburst remained
fairly intact during the liftoff, while most of the mass lost in the
last outburst disintegrated into dust very soon.  This scenario is
supported by the detection of a cloud of material, found to have been
ejected in early October,~at a time that closely correlates with the time
of the last outburst and the birth of companion C.  All six companions
belong to a group of very brittle fragments, which explains their short
lifetimes and elongated shape.\\

I thank Carl W.\ Hergenrother for his unpublished measurement of the
position of a suspected companion --- that turned out to be fragment D ---
in the images of comet 168P he took with the VATT reflector on November
17.  I also thank him for reading and commenting on the manuscript of
this paper.  Finally, I thank Dr. Daniel W.\ E.\ Green for his editorial
work.  This research was carried out at the Jet Propulsion Laboratory,
Cali\-fornia Institute of Technology, under contract with the National
Aeronautics and Space Administration.\\[-0.1cm]
\begin{center}
{\footnotesize REFERENCES}
\end{center}
%
%
{\footnotesize
\parbox{8.6cm}{Abetti, A. (1883). Osservazioni della cometa Pons-Brooks,
 {\it Astron. {\hspace*{0.3cm}}Nachr.\/} {\bf 107}, 223--228.}\\[0.05cm]
\parbox{8.6cm}{A'Hearn, M. F.: D. G. Schleicher; P. D. Feldman; R. L. Millis;
{\hspace*{0.3cm}}and D. T. Thompson (1984). Comet Bowell 1980b, {\it Astron.
J.\/} {\hspace*{0.3cm}}{\bf 89}, 579--591.}\\[0.05cm]
\parbox{8.6cm}{A'Hearn, M. F.: R. L. Millis; D. G. Schleicher; D. J. Osip;
and {\hspace*{0.3cm}}P. V. Birch (1995). The ensemble properties of comets:\
Results {\hspace*{0.3cm}}from narrowband photometry of 85 comets, 1976--1992,
{\it Icarus\/} {\hspace*{0.3cm}}{\bf 118}, 223--270.}\\[0.05cm]
\parbox{8.6cm}{Beyer, M. (1962). Physische Beobachtungen von Kometen. XII,
{\hspace*{0.3cm}}{\it Astron.  Nachr.\/} {\bf 286}, 219--240.}\\[0.05cm]
\parbox{8.6cm}{Bobrovnikoff, N. T. (1931). Halley's Comet in its apparition
of {\hspace*{0.3cm}}1909--1911, {\it Publ. Lick Obs.\/} {\bf 17},
305--482.}\\[0.05cm]
\parbox{8.6cm}{Bobrovnikoff, N. T. (1932). On the phenomena of halos in
comets, {\hspace*{0.3cm}}{\it Publ.  Astron. Soc. Pacific\/} {\bf 44},
296--307.}\\[0.05cm]
\parbox{8.6cm}{Bobrovnikoff, N. T. (1943). The Periodic Comet Holmes
(1892 {\hspace*{0.3cm}}III), {\it Pop.  Astron.\/} {\bf 51}, 542--551.}\\[0.05cm]
\parbox{8.6cm}{Campbell, W. W. (1893). The spectrum of Holmes' Comet, {\it
Publ. {\hspace*{0.3cm}}Astron. Soc.  Pacific\/} {\bf 5}, 99--100.}\\[0.05cm]
\parbox{8.6cm}{Chandler, S. C., Jr. (1883). On the outburst in the light of
the {\hspace*{0.3cm}}comet Pons-Brooks Sept. 21--23, {\it Astron.  Nachr.\/}
{\bf 107}, 131--134.}\\[0.05cm]
\parbox{8.6cm}{Crovisier, J., D. Bockel\'ee-Morvan; E. G\'erard; H. Rauer;
N.~Biver; {\hspace*{0.3cm}}P. Colom; and L. Jorda (1996). What happened to
comet {\hspace*{0.3cm}}73P/Schwassmann-Wachmann 3?, {\it Astron.\ Astrophys.\/}
{\bf 310}, {\hspace*{0.3cm}}L17--L20.}\\[0.05cm]
\parbox{8.6cm}{Gonzalez, J. J. (2012). Comet 168P/Hergenrother, {\it Centr.
Bureau {\hspace*{0.3cm}}Electr. Tel.\/} 3257.}\\[0.05cm]
\parbox{8.6cm}{Green, D. W. E., ed. (1995). Comet 41P/Tuttle-Giacobini-Kres\'ak,
{\hspace*{0.3cm}}{\it IAU Circ.\/} 6207.}\\[0.05cm]
\parbox{8.6cm}{Green, D. W. E., ed. (1997a). Astrometry of comets, in {\it
The ICQ {\hspace*{0.3cm}}Guide to Observing Comets\/}, Smithsonian Astrophysical
Obser- {\hspace*{0.3cm}}vatory, Cambridge, Mass., pp. 151--157.}\\[0.05cm]
\parbox{8.6cm}{Green, D. W. E., ed. (1997b). CCD observations, in {\it The
ICQ {\hspace*{0.3cm}}Guide to Observing Comets\/}, Smithsonian Astrophysical
Obser- {\hspace*{0.3cm}}vatory, Cambridge, Mass., pp. 94--104.}\\[0.05cm]
\parbox{8.6cm}{Green, D. W. E., ed. (2012). Comet 168P/Hergenrother, {\it
Centr. {\hspace*{0.3cm}}Bureau Electr. Tel.\/} 3257.}\\[0.05cm]
\parbox{8.6cm}{Hergenrother, C. W. (2012a). Comet 168P/Hergenrother, {\it
Centr. {\hspace*{0.3cm}}Bureau Electr. Tel.\/} 3295.}\\[0.05cm]
\parbox{8.6cm}{Hergenrother, C. W. (2012b). Comet 168P/Hergenrother, {\it
Centr. {\hspace*{0.3cm}}Bureau Electr. Tel.\/} 3318.}\\[0.05cm]
\parbox{8.6cm}{Herschel, J. F. W. (1847). {\it Results of Astronomical
Observations {\hspace*{0.3cm}}Made During the Years 1834, 5, 6, 7, 8, at
the Cape of Good {\hspace*{0.3cm}}Hope\/}, Chapter V, pp. 393--413.  Smith,
Elder \& Co., Cornhill. {\hspace*{0.3cm}}London, U.K.}\\[0.05cm]
\parbox{8.6cm}{Jehin, E.; H. Boehnhardt; Z. Sekanina; X. Bonfils; O. Sch\"{u}tz;
J.- {\hspace*{0.3cm}}L. Beuzit; M. Billeres; G. J. Garradd; P. Leisy; and F.
Marchis {\hspace*{0.3cm}}(2002). Split comet C/2001 A2 (LINEAR), {\it Earth
Moon Plan.\/} {\hspace*{0.3cm}}{\bf 90}, 147--151.}\\[0.05cm]
\parbox{8.6cm}{Jewitt, D. (1990). The persistent coma of Comet
P/Schwassmann- {\hspace*{0.3cm}}Wachmann 1, {\it Astrophys. J.\/} {\bf 351},
277--286.}\\[0.05cm]
\parbox{8.6cm}{Kammermann, A. (1893). Beobachtungen des Cometen 1892 III
{\hspace*{0.3cm}}(Holmes), {\it Astron. Nachr.\/} {\bf 132}, 61--62.}\\[0.05cm]
\parbox{8.6cm}{Kidger, M. R. (2002). Spanish monitoring of comets: Making
sense {\hspace*{0.3cm}}of amateur photometric data, {\it Earth Moon Plan.\/}
{\bf 90}, 259--268.}\\[0.05cm]
\parbox{8.6cm}{Kres\'ak, L'. (1974). The outbursts of periodic comet
Tuttle- {\hspace*{0.3cm}}Giacobini-Kres\'ak, {\it Bull. Astron. Inst. Czech.\/}
{\bf 25}, 293--304.}\\[0.05cm]
\parbox{8.6cm}{Li, J.; D. Jewitt; J. M. Clover; and B. V. Jackson (2011).
Out- {\hspace*{0.3cm}}burst of Comet 17P/Holmes observed with the Solar Mass
Ejec- {\hspace*{0.3cm}}tion Imager, {\it Astrophys.  J.\/} {\bf 728},
31--39.}\\[0.05cm]
\parbox{8.6cm}{Meech, K. J.; M. J. S. Belton; B. E. A. Mueller; M.~W.~Dicksion;
{\hspace*{0.3cm}}and H. R. Li (1993). Nucleus properties of
P/Schwassmann- {\hspace*{0.3cm}}Wachmann 1. {\it Astron.  J.\/} {\bf 106},
1222--1236.}\\[0.05cm]
\parbox{8.6cm}{Moreno, F. (2009). The dust environment of Comet
29P/Schwass- {\hspace*{0.3cm}}mann-Wachmann 1 from dust tail modeling of
2004~near-{\linebreak} {\hspace*{0.3cm}}perihelion observations,
{\it Astrophys. J. Suppl. Ser.\/} {\bf 183}, 33--45.}\\[0.05cm]
\parbox{8.6cm}{M\"{u}ller, G. (1884a). \"{U}ber einen zweiten merkw\"{u}rdigen
Lichtaus- {\hspace*{0.3cm}}bruch an dem Cometen Pons-Brooks 1884~Jan.\,1,
{\it Astron. Nachr.\/} {\hspace*{0.3cm}}{\bf 107}, 381--384.}\\[-0.15cm]
\clearpage
\noindent
\parbox{8.6cm}{M\"{u}ller, G. (1884b). Photometrische Beobachtungen des
Cometen {\hspace*{0.3cm}}Pons-Brooks, {\it Astron. Nachr.\/} {\bf 108},
161--170.}\\[0.04cm]
\parbox{8.6cm}{Palisa, J. (1893). Ueber eine pl\"{o}tzliche Aenderung im
Aussehen {\hspace*{0.3cm}}des Cometen 1892 III (Holmes), {\it Astron.
Nachr.\/} {\bf 132}, 15--16, {\hspace*{0.3cm}}31--32.}\\[0.04cm]
\parbox{8.6cm}{Pickering, E. C.; A. Searle; and O. C. Wendell (1900).
Observa- {\hspace*{0.3cm}}tions of comets, {\it Ann. Harv. Coll. Obs.\/}
{\bf 33}, 149--158.}\\[0.04cm]
\parbox{8.6cm}{Schiaparelli, G. V. (1883). Osservazioni della cometa
Pons-Brooks, {\hspace*{0.3cm}}{\it Astron.  Nachr.\/} {\bf 107},
139--144.}\\[0.04cm]
\parbox{8.6cm}{Schleicher, D. G. (2012). Comet 168P/Hergenrother, {\it
Centr. Bu- {\hspace*{0.3cm}}reau Electr. Tel.\/} 3257.}\\[0.04cm]
\parbox{8.6cm}{Schlosser, W.; R. Schulz; and P. Koczet (1986). The cyan
shells {\hspace*{0.3cm}}of comet P/Halley, in {\it Exploration of Halley's
Comet\/}, Proc. 20th {\hspace*{0.3cm}}ESLAB Symp., ESA SP-250, B. Battrick,
E. J. Rolfe, and R. {\hspace*{0.3cm}}Reinhard, eds., ESTEC, Noordwijk,
vol. 3, pp. 495--498.}\\[0.04cm]
\parbox{8.6cm}{Schulz, R., and W. Schlosser (1990). CN jets as progenitors
of CN {\hspace*{0.3cm}}shells in the coma of Comet P/Halley, in {\it
Formation of Stars and {\hspace*{0.3cm}}Planets, and the Evolution of the
Solar System\/}, ESA SP-315, {\hspace*{0.3cm}}B. Battrick, ed., ESTEC,
Noordwijk, pp. 121--125.}\\[0.04cm]
\parbox{8.6cm}{Sekanina, Z. (1977). Relative motions of fragments of the
split {\hspace*{0.3cm}}comets. I. A new approach, {\it Icarus\/} {\bf 30},
574--594.}\\[0.04cm]
\parbox{8.6cm}{Sekanina, Z. (1978). Relative motions of fragments of the
split {\hspace*{0.3cm}}comets.  II. Separation velocities and differential
decelerations {\hspace*{0.3cm}}for extensively observed comets, {\it
Icarus\/} {\bf 33}, 173--185.}\\[0.04cm]
\parbox{8.6cm}{Sekanina, Z. (1982). The problem of split comets in review,
in {\hspace*{0.3cm}}{\it Comets\/}, L. L. Wilkening, ed., University of
Arizona, Tucson, {\hspace*{0.3cm}}pp. 251--287.}\\[0.04cm]
\parbox{8.6cm}{Sekanina, Z. (1990). Gas and dust emission from comets and
life {\hspace*{0.3cm}}spans of active areas on their rotating nuclei, {\it
Astron. J.\/} {\bf 100}, {\hspace*{0.3cm}}1293--1314, 1389--1391.}\\[0.04cm]
\parbox{8.6cm}{Sekanina, Z. (1993). Computer simulation of the evolution of
dust {\hspace*{0.3cm}}coma morphology in an outburst: P/Schwassmann-Wachmann
{\hspace*{0.3cm}}1, in {\it On the Activity of Distant Comets\/}, W. F.
Huebner, H. {\hspace*{0.3cm}}U. Keller, D. Jewitt, J. Klinger, and R. West,
eds., Southwest {\hspace*{0.3cm}}Research Institute, San Antonio, TX, pp.
166--181.}\\[0.04cm]
\parbox{8.6cm}{Sekanina, Z. (2005). Comet 73P/Schwassmann-Wachmann: Nu-
{\hspace*{0.3cm}}cleus fragmentation, its light-curve signature, and close
approach {\hspace*{0.3cm}}to earth in 2006, {\it Int.  Comet Quart.\/}
{\bf 27}, 225--240.}\\[0.04cm]
\parbox{8.6cm}{Sekanina, Z. (2008a). Exploding comet 17P/Holmes, {\it Int.
Comet {\hspace*{0.3cm}}Quart.\/} {\bf 30}, 3--28.}\\[0.03cm]
\parbox{8.6cm}{Sekanina, Z. (2008b). On a forgotten 1836 explosion from
Hal- {\hspace*{0.3cm}}ley's comet, reminiscent of 17P/Holmes' outbursts,
{\it Int. Comet {\hspace*{0.3cm}}Quart.\/} {\bf 30}, 63--74.}\\[0.03cm]
\parbox{8.6cm}{Sekanina, Z. (2012a). Comet 168P/Hergenrother, {\it Centr.
Bureau {\hspace*{0.3cm}}Electr. Tel.\/} 3295.}\\[0.03cm]
\parbox{8.6cm}{Sekanina, Z. (2012b). Comet 168P/Hergenrother, {\it Centr.
Bureau {\hspace*{0.3cm}}Electr. Tel.\/} 3318.}\\[-0.45cm]
\parbox{8.6cm}{Sekanina, Z.; and P. W. Chodas (2002). Common origin of two
{\hspace*{0.3cm}}major sungrazing comets, {\it Astrophys. J.\/} {\bf 581},
760--769.}\\[0.08cm]
\parbox{8.6cm}{Sekanina, Z., and P. W. Chodas (2012). Comet C/2011 W3
{\hspace*{0.3cm}}\mbox{(Lovejoy)}: Orbit determination, outbursts,
disintegration of nu- {\hspace*{0.3cm}}cleus, dust-tail morphology, and
relationship to new cluster of {\hspace*{0.3cm}}bright sungrazers, {\it
Astrophys. J.\/} {\bf 757}, 127 (33pp).}\\[0.1cm]
\parbox{8.6cm}{Sekanina, Z.; E. Jehin; H. Boehnhardt; X. Bonfils; O. Schuetz;
and {\hspace*{0.3cm}}D. Thomas (2002). Recurring outbursts and nuclear
fragmenta- {\hspace*{0.3cm}}tion of comet C/2001~A2 (LINEAR), {\it Astrophys.
J.\/} {\bf 572}, 679--{\hspace*{0.3cm}}684.}\\[0.1cm]
\parbox{8.6cm}{Sostero, G.; N. Howes; A. Tripp; P. Phelps; and E. Guido
(2012). {\hspace*{0.3cm}}Comet 168P/Hergenrother, {\it Centr. Bureau
Electr. Tel.\/} 3295.}\\[0.09cm]
\parbox{8.6cm}{Spahr, T. B.; G. V. Williams; S. Nakano; and A. Doppler,
eds. {\hspace*{0.3cm}}(2012). Comet 168P/Hergenrother, {\it Minor Planet
Circ.\/} 80422--{\hspace*{0.3cm}}80429.}\\[0.09cm]
\parbox{8.6cm}{Stansberry, J. A.; J. Van Cleve; W. T. Reach; D. P.
Cruikshank; {\hspace*{0.3cm}}J. P. Enery; Y. R. Fernandez; V. S. Meadows;
K. Y. L. Su; K. {\hspace*{0.3cm}}Misselt; G. H. Rieke; E. T. Young; M. W.
Werner; C. W. Engel- {\hspace*{0.3cm}}bracht; K. D. Gordon; D. C. Hines;
D. M. Kelly; J. E. Morrison; {\hspace*{0.3cm}}and J. Muzerolle (2004).
Spitzer observations of the dust coma {\hspace*{0.3cm}}and nucleus of
29P/Schwassmann-Wachmann 1, {\it Astrophys. J. {\hspace*{0.3cm}}Suppl.
Ser.\/} {\bf 154}, 463--468.}\\[0.09cm]
\parbox{8.6cm}{Stevenson, R. A.; J. M. Bauer; J. R. Masiero; and A. K.
Mainzer {\hspace*{0.3cm}}(2012).  Comet 168P/Hergenrother, {\it Centr.
Bureau. Electr. Tel.\/} {\hspace*{0.3cm}}3295.}\\[0.09cm]
\parbox{8.6cm}{Struve, H. (1884). Beobachtungen des Cometen 1884 I (Pons
1812) {\hspace*{0.3cm}}am 15 z\"{o}ll. Refractor in Pulkowa, {\it Astron.
Nachr.\/} {\bf 109}, 369--386.}\\[0.09cm]
\parbox{8.6cm}{Swings, J.-P.; and J.-M. Vreux (1973). Observations
spectro- {\hspace*{0.3cm}}graphiques de la com\`ete 1973b imm\'ediatement
apr\`es son vio- {\hspace*{0.3cm}}lent sursaut de brillance de juillet
1973, {\it Bull. Soc. Roy. Sci. {\hspace*{0.3cm}}Li\`ege\/} {\bf 42},
593--597.}\\

\vspace{0.08cm}
\noindent
\parbox{8.6cm}{Trigo-Rodriguez, J. M.; E. Garcia-Melendo; B. J. R.
Davidsson; {\hspace*{0.3cm}}A. S\'anchez; D. Rodriguez; J. Lacruz;
J. A. De los Reyes; and S. {\hspace*{0.3cm}}Pastor (2008). Outburst
activity in comets. I, {\it Astron.  Astro- {\hspace*{0.3cm}}phys.\/}
{\bf 485}, 599--606.}\\[0.09cm]
\parbox{8.6cm}{Trigo-Rodriguez, J. M.; D. A. Garcia-Hern\'andez; A.
S\'anchez; J. {\hspace*{0.3cm}}Lacruz; B. J. R. Davidsson; D. Rodriguez;
S. Pastor; and J. A. {\hspace*{0.3cm}}de los Reyes (2010).  Outburst
activity in comets. II, {\it Mon.  Not. {\hspace*{0.3cm}}Roy. Astron.
Soc.\/} {\bf 409}, 1682--1690.}\\[0.08cm]
\parbox{8.6cm}{Vogel, H. C. (1884). Einige Beobachtungen \"{u}ber den
Cometen {\hspace*{0.3cm}}Pons-Brooks, insbesondere \"{u}ber das Spectrum
desselben, {\it As- {\hspace*{0.3cm}}tron. Nachr.\/} {\bf 108},
21--26.}\\[0.07cm]
\parbox{8.6cm}{Vogel, H. C. (1893). Ueber das Spectrum des Cometen
1892 III {\hspace*{0.3cm}}(Holmes), {\it Astron. Nachr.\/} {\bf 131},
373--374.}\\[0.07cm]
\parbox{8.6cm}{Whipple, F. L. (1980). Rotation and outbursts of comet
{\hspace*{0.3cm}}P/Schwassmann-Wachmann 1, {\it Astron. J.\/} {\bf 85},
305--313.}\\[-1cm]
\end{document}